\documentclass[preprint,12pt]{elsarticle}

\usepackage{amssymb}
\usepackage{dcolumn}% Align table columns on decimal point
\usepackage{bm}% bold math
\usepackage{braket}
\usepackage{amsmath}
\usepackage{color}
\usepackage[outdir=./]{epstopdf}

\usepackage{lineno}
\newcommand*\patchAmsMathEnvironmentForLineno[1]{
  \expandafter\let\csname old#1\expandafter\endcsname\csname #1\endcsname
  \expandafter\let\csname oldend#1\expandafter\endcsname\csname end#1\endcsname
  \renewenvironment{#1}
     {\linenomath\csname old#1\endcsname}
     {\csname oldend#1\endcsname\endlinenomath}}
\newcommand*\patchBothAmsMathEnvironmentsForLineno[1]{
  \patchAmsMathEnvironmentForLineno{#1}
  \patchAmsMathEnvironmentForLineno{#1*}}
\AtBeginDocument{
\patchBothAmsMathEnvironmentsForLineno{equation}
\patchBothAmsMathEnvironmentsForLineno{align}
\patchBothAmsMathEnvironmentsForLineno{flalign}
\patchBothAmsMathEnvironmentsForLineno{alignat}
\patchBothAmsMathEnvironmentsForLineno{gather}
\patchBothAmsMathEnvironmentsForLineno{multline}
}
\makeatletter
\def\mojiparline#1{
\newcounter{mpl}
\setcounter{mpl}{#1}
\@tempdima=\linewidth
\advance\@tempdima by-\value{mpl}zw
\addtocounter{mpl}{-1}
\divide\@tempdima by \value{mpl}
\advance\kanjiskip by\@tempdima
\advance\parindent by\@tempdima
}
\makeatother

\newcommand{\argmax}{\mathop{\rm arg~max}\limits}

\journal{Physica D}

\begin{document}

\begin{frontmatter}

%% Title, authors and addresses

%% use the tnoteref command within \title for footnotes;
%% use the tnotetext command for theassociated footnote;
%% use the fnref command within \author or \address for footnotes;
%% use the fntext command for theassociated footnote;
%% use the corref command within \author for corresponding author footnotes;
%% use the cortext command for theassociated footnote;
%% use the ead command for the email address,
%% and the form \ead[url] for the home page:
%% \title{Title\tnoteref{label1}}
%% \tnotetext[label1]{}
%% \author{Name\corref{cor1}\fnref{label2}}
%% \ead{email address}
%% \ead[url]{home page}
%% \fntext[label2]{}
%% \cortext[cor1]{}
%% \affiliation{organization={},
%%             addressline={},
%%             city={},
%%             postcode={},
%%             state={},
%%             country={}}
%% \fntext[label3]{}

\title{Procedure to Reveal the Mechanism of Pattern Formation Process by Topological Data Analysis}

%% use optional labels to link authors explicitly to addresses:
%% \author[label1,label2]{}
%% \affiliation[label1]{organization={},
%%             addressline={},
%%             city={},
%%             postcode={},
%%             state={},
%%             country={}}
%%
%% \affiliation[label2]{organization={},
%%             addressline={},
%%             city={},
%%             postcode={},
%%             state={},
%%             country={}}

\author[inst1]{Yoh-ichi Mototake}
\affiliation[inst1]{organization={Graduate School of Social Data Science, Hitotsubashi University},%Department and Organization
            addressline={2-1, Naka}, 
            city={Kunitachi},
            postcode={186-8601}, 
            state={Tokyo},
            country={Japan}}

\author[inst2]{Masaichiro Mizumaki}
\affiliation[inst2]{organization={Graduate School of Science and Technology, Kumamoto University},%Department and Organization
            addressline={2-39-1 Kurokami, Chuo-ku}, 
            city={Kumamoto},
            postcode={860-8555}, 
            state={Kumamoto},
            country={Japan}}

\author[inst3,inst4]{Kazue Kudo}
\affiliation[inst3]{organization={Department of Computer Science, Ochanomizu University},%Department and Organization
            addressline={2-1-1 Otsuka}, 
            city={Bunkyo-ku},
            postcode={112-8610}, 
            state={Tokyo},
            country={Japan}}
\affiliation[inst4]{organization={Department of Computer and Mathematical Sciences, Tohoku University},%Department and Organization
            addressline={6-3-09 Aoba, Aramaki-aza Aoba-ku}, 
            city={Sendai},
            postcode={980-8579}, 
            state={Miyagi},
            country={Japan}}

\author[inst5]{Kenji Fukumizu}
\affiliation[inst5]{organization={The Institute of Statistical Mathematics},%Department and Organization
            addressline={10-3 Midori-cho}, 
            city={Tachikawa},
            postcode={190-8562}, 
            state={Tokyo},
            country={Japan}}

\begin{abstract}
%% Text of abstract
Topological data analysis (TDA) is a versatile tool that can be used to extract scientific knowledge from complex pattern formation processes. 
However, the physics correspondence between the features obtained from TDA and pattern dynamics does not agree one-to-one, and the physical interpretation of the TDA features needs to be set appropriately according to the phenomenon to be analyzed. 
In this study, we propose an analytical procedure to physically interpret pattern dynamics through TDA and machine learning techniques. 
The proposed procedure was applied to the process of magnetic domain pattern formation to quantify non-trivial domain pattern classifications and reveal the nature of the underlying dynamics. 
On the basis of these findings, we also propose a candidate reduction model to understand the nature of magnetic domain formation. 
%In this paper, we demonstrate the effectiveness of TDA in understanding magnetic domain formation in magnetic thin films. Our findings show that the persistence diagrams (PDs) corresponding to different magnetic domain patterns contain information on both the domain structure and interdomain fluctuations. We also propose a quantitative classification of these patterns based on their PDs and a candidate reduction model to describe their mechanics.
\end{abstract}

%%Graphical abstract
%\begin{graphicalabstract}
%\includegraphics[width=\linewidth]{./fig3.pdf}
%\end{graphicalabstract}

%%Research highlights
%\begin{highlights}
%\item We propose an analytical procedure to physically interpret pattern dynamics through topological data analysis (TDA) and machine learning techniques. 
%\item The proposed procedure was applied to magnetic domain pattern formation, and to quantify non-trivial domain pattern classifications and reveal the nature of the underlying dynamics.
%\item Pattern classification was statistically understood by using inverse analysis of topological features.
%\item It was also confirmed that the intensity fluctuations in magnetic moment of the magnetic domain structures contain useful information about the system.
%\item Based on the quantified properties of the dynamics, a reduced model of the pattern dynamics was developed.
%\item Based on the developed reduced model, the mechanism underlying pattern classification based on TDA was elucidated.
%\end{highlights}

\begin{keyword}
%% keywords here, in the form: keyword \sep keyword
topological data analysis \sep pattern dynamics \sep magnetic domain pattern \sep machine learning \sep interpretability of pattern dynamics
%% PACS codes here, in the form: \PACS code \sep code
%\PACS 89.75.Kd
%% MSC codes here, in the form: \MSC code \sep code
%% or \MSC[2008] code \sep code (2000 is the default)
%\MSC 62-07 %\sep 1111
\end{keyword}

\end{frontmatter}

%% \linenumbers

%% main text
\section{Introduction}
The formation of nonuniform and nonperiodic structures through pattern formation processes with long-range interactions is crucial for comprehending the physical characteristics of complex systems in nature, such as ferromagnetic materials, structural materials, and chemical reactions.
These processes are not typically based on the smallest governing elements, such as spins, atoms, or molecules, but rather on spatially coarse-grained features representing microdomains, such as magnetic moment, reflection ratio, or density.
Mathematical models, like the time-dependent Ginzburg--Landau (TDGL) equation~\cite{jagla2004numerical, kudo2007magnetic}, phase-field model~\cite{chen2002phase}, and reaction-diffusion model~\cite{gray1983autocatalytic}, are commonly used to model pattern formation processes and are based on continuum approximation models.
The measurement or modeling of pattern formation dynamics is essentially done on a state space, or the pixel space of a grayscale image.
However, the nonuniform and nonperiodic structure of a grayscale image makes it challenging to extract meaningful features to characterize pattern formation processes, as traditional methods like statistical values for random phenomena or Fourier basis for periodic structures are not applicable.
Therefore, determining effective methods for extracting and analyzing pattern formation phenomena from grayscale images is a significant challenge in the field of pattern formation.\par

An important challenge is to determine how to extract effective features from grayscale image data with the aforementioned characteristics. 
In conventional studies~\cite{kudo2007field, guiu2012characterizing, nahas2020topology}, grayscale images are often separated into domains and other regions by binarizing them using a certain threshold, and pattern formation phenomena are understood by the geometrical features of the domain structure, such as the curvature, area of the domain, and surface area. 
On the basis of Gauss--Bonnet theorem, the curvature is associated with a topological feature called the Euler characteristic. 
The area of the domain, the boundary area, and the Euler characteristic constitute the free energy that governs certain pattern formation dynamics~\cite{konig2004morphological}. 
However, when binarizing a grayscale image, information such as fluctuations in the domain may be lost. 
From this perspective, changes in the topological features of the domain structure at different binarization thresholds have been investigated in a previous study~\cite{mecke1996morphological}. 
The study suggested that the change in the Euler characteristic is a particularly promising feature that behaves as an order parameter that characterizes a certain pattern phase. 
In the previous studies introduced thus far, the sum or average of the geometrical structures of all domains in the entire grayscale image is considered as a feature. 
In a nonuniform and nonperiodic pattern formation process, each isolated geometrical structures in the grayscale image are able to have different topological features. 
Therefore, to determine the properties of nonuniform and nonperiodic pattern formation processes, we need geometrical features that can represent the changes in the topological features of each isolated domain according to a threshold value. \par
To characterize and understand the pattern formation dynamics with long-range interaction that form nonuniform and nonperiodic structures presented as grayscale images, we employ a topological data analysis (TDA)~\cite{carlsson2009topology, edelsbrunner2014short} method to determine the changes in the topological features of each isolated domain according to the threshold value. 
TDA is a general term for data analysis methods that use the concept of topology. 
Of the methods comprising TDA, analysis methods based on a persistent homology group~\cite{carlsson2009topology, edelsbrunner2014short} achieve the desired feature extraction. 
Persistent homology is a mathematical structure that describes the birth and death of ``hole'' structures due to changes in a single parameter, such as the binarization threshold of grayscale images, which characterizes the geometrical structure. 
A hole is a ring structure in the one-dimensional case and a cavity structure in the two-dimensional case. 
Specifically,  the number of holes corresponding to the Betti number is a homotopy invariant, that is, invariant to continuous changes of the geometrical structure, and the Betti number constitutes the Euler characteristic. 
Persistent homology is a topological invariant that is an extension of the Betti number to describe the birth and death of individual hole structures as the binarization threshold value increases. 
Thus, feature extraction based on persistent homology can satisfy the conditions for obtaining effective features described in the previous paragraph. 
Persistent homology has already been applied to pattern formation data provided by grayscale images. 
D{\l}otko and Wanner applied TDA to the pattern formation dynamics of a Cahn--Hilliard system of diffusion equations and demonstrated that it is possible to estimate the total mass, which is the statistics representing the property of the pattern, and the time of a snapshot image from the extracted features~\cite{dlotko2016topological}.
Furthermore, Calcina and Gameiro applied TDA to the patterns generated using the Ginzburg--Landau equation and reported that the parameters of the simulation model can be estimated from formed patterns with high accuracy~\cite{calcina2021parameter}. 
The results of these previous studies conducted to model parameter estimation indicate that the features extracted from persistent homology retain rich information about pattern formation dynamics. 
Therefore, physical discovery, such as elucidating unquantified pattern states or mechanisms of pattern formation dynamics, can be achieved by detailed analysis and interpretation of the features of pattern formation dynamics extracted from persistent homology. 
From the viewpoint of obtaining novel scientific knowledge about pattern formation dynamics, analysis methods based on persistent homology have the following two advantages. 
First, persistent homology can be inversely analyzed such that from the extracted persistent homology structure, the representative locations of the corresponding holes in the original space~\cite{obayashi2018persistence, vandaele2020topological}, the geometrical structure in the original space corresponding to the topological features~\cite{dey2011optimal}, and the entire original pattern can be estimated~\cite{gameiro2016continuation}. 
Therefore, features obtained from persistent homology are highly interpretable. 
This is useful in discussing the underlying physical mechanisms on the basis of extracted features. 
Second, persistent homology satisfies the stability theorem~\cite{cohen2007stability}, which states that when the spatial structure of data changes slightly, the persistent homology also changes only slightly. 
This is a useful property of persistent homology in discussing phase-transition-like phenomena of geometrical structures, ensuring that false detection of phase transition does not occur. 
Thus, a analysis methods based on persistent homology is useful for revealing the mechanism of nonuniform and nonperiodic pattern formation processes provided as grayscale images.\par
The purpose of this study is to propose an analysis procedure based on persistent homology which can facilitate scientific discoveries in complex pattern formation dynamics provided as grayscale image data.  
We verify it by applying the procedure to the process of magnetic domain pattern formation under a rapidly sweeping external magnetic field. 
A magnetic domain pattern, which is a spatial pattern formed by a macroscopic magnetic field, is an important feature that is closely related to the mechanism of coercivity~\cite{hubert2008magnetic}. 
For instance, the nonperiodic structure of magnetic domain patterns generated by spatial inhomogeneities, such as defects, impurities, steps, and strains, in the crystal structure of magnetic materials is closely related to the mechanism by which high coercivity is exhibited. 
As described above, the ``shape'' of a magnetic domain pattern is intrinsically important to the properties of ferromagnets; thus, it is important to clarify the relationship between the shape of a magnetic domain pattern and the physical properties~\cite{hubert2008magnetic}. 
Measurement and simulation data of magnetic domain patterns often consist of a set of average magnetic moments in a microdomain. 
This signifies that a grayscale pixel image is the state space in which the dynamics are discussed. 
In conventional analysis, the grayscale images representing the magnetic domain pattern are binarized, and geometrical features, such as the number of isolated domains or the domain length, are extracted for analysis~\cite{kudo2007magnetic, kudo2007field}. 
This analysis makes it possible to classify magnetic domain patterns and elucidate some of the physical mechanisms underlying their formation. 
The introduction of TDA using persistent homology, which does not assume binarization, can elucidate the physical mechanism of more elaborate or extensive phenomena of magnetic domain pattern formation. 
In this study, we analyze time-series image data of the magnetic domain pattern formation generated by a simulation based on the TDGL equation. 
Because the TDGL equation is closely related to a wide class of pattern formation processes, confirming the usefulness of persistent homology in the TDGL equation will indicate that persistent homology is also useful for elucidating many other pattern formation dynamics.\par
The remainder of this paper is organized as follows.
First, we describe in  Sec.~\ref{sec_tdgl} the TDGL equation that simulates the process of magnetic domain pattern formation to be analyzed. 
Next, persistent homology and proposed analytical procedure to extract new knowledge from the dynamics data of magnetic domain patterns using persistent homology are described in Sec.~\ref{sec_method}. 
These analysis results are presented in Secs.~\ref{sec_result1} and \ref{sec_result2}. 
Then, on the basis of the results of these analyses, the physical mechanisms of magnetic domain pattern formation and advantages of features based on TDA are discussed in Sec.~\ref{sec_discussion}. 
A summary of the study is provided in Sec.~\ref{summary}.
Table~\ref{tbl0} also summarizes a list of abbreviations and symbols frequently used in this paper for easy readability.

\begin{table}[b]
\caption{\label{tbl0}%
Summary of notations.}

\scalebox{0.6}[0.6]{
\begin{tabular}{c|l}
\hline
Abbreviation and symbols & \multicolumn{1}{c}{Description}\\ \hline\hline
TDGL equation& 
Time-dependent Ginzburg--Landau equation
\\ \hline
TDA& 
Topological data analysis
\\ \hline
%PH& 
%abbreviation of persistent homology
%\\ \hline
PD& 
Persistence diagram\\ \hline
YIG & 
Yttrium iron garnet\\ \hline
${\rm PD_{all}}$ & Feature vector of a persistence diagram\\ \hline
${\rm PD_{long}}$ & Subspace of feature vector of a persistence diagram around long-life region \\ \hline
${\rm PD_{short}}$ & Subspace of feature vector of a persistence diagram around short-life region \\ \hline
RMSE & 
Root mean squared error \\ \hline
PCA & 
Principal component analysis\\ \hline
$\phi$ & 
$z$ component of magnetic moment\\ \hline
$\alpha$ & 
Coefficient of anisotropic energy\\ \hline
$v$ & 
Sweep rate of external magnetic field h(t) \\ \hline
$G_i$ & 
Grayscale image $\textcolor{black}{i}$. \\ \hline
${\bf v}_i(\sigma,\textcolor{black}{M})$ & 
Vector data of a grayscale image $G_i$ obtained using vectorized parameters $\sigma$ and $\textcolor{black}{M}$. \\ \hline
\end{tabular}
}
\end{table}

\section{Pattern Formation Model: TDGL equation}
\label{sec_tdgl}
There are a large number of physical systems that display the pattern formation process. 
Yttrium iron garnet (YIG) thin films, which are among the most prominent materials used in the research of spin dynamics in thin films~\cite{hauser2016yttrium}, can show complex magnetic domain patterns formed by the macroscopic magnetic field. 
For example, depending on the sweeping rate of the external magnetic field, various nonuniform and nonperiodic patterns can be formed under a zero magnetic field. 
Spins in certain types of YIG thin film have a strong uniaxial anisotropy in the normal direction of the film. 
Therefore, in the limit where the magnetic anisotropy of the perpendicular component of the film surface is strong and the elements of other directions are zero, the process of the magnetic domain pattern formation in the YIG thin film is sometimes modeled on the scalar field $\phi\left(\bm{r}\right):\: \mathbf{R}^2\rightarrow \mathbf{R}$, $\bm{r}=\left(x,y\right)$, which is composed of the average of spin-perpendicular components of small regions arranged in a lattice~\cite{jagla2004numerical, kudo2007magnetic}. 
In this continuum approximation model, the pattern formation process is described on the basis of the principle that the pattern is formed along the gradient of the energy functional composed of the mean field $\phi\left(\bm{r}\right)$. 
In this study, on the basis of the work of Kudo et al.~\cite{kudo2007magnetic}, the time-series data of the magnetic domain pattern formation in a fast sweeping external magnetic field are generated using the TDGL equation. 
The variation of the pattern with the sweeping speed of the external magnetic field has been observed experimentally~\cite{kudo2007magnetic}. 
Understanding the pattern formation process with such complex interactions with the environment can lead to an understanding of the properties of YIG thin films in various application situations. 
The TDGL equation employed for pattern formation data generation is formulated as follows: 
\begin{eqnarray}
\label{eq_variation}
\frac{\partial\phi(\bm{r})}{\partial t} &=& -\frac{\delta H(\phi(\bm{r}))}{\delta \phi(\bm{r})},\\
\label{energy}
H(\phi(\bm{r})) &=& \alpha \int d\bm{r} \lambda(\bm{r}) \left(\frac{\phi(\bm{r})^4}{4} -\frac{\phi(\bm{r})^2}{2} \right) + \beta \int d\bm{r} \frac{\left|\nabla \phi(\bm{r})\right|^2}{2}\nonumber\\
&\:&+ \gamma \int_{\left|\bm{r} - \bm{r}'\right| > \textcolor{black}{r_0}} d\bm{r} d\bm{r}' \frac{\phi(\bm{r})\phi(\bm{r}')}{\left|\bm{r} - \bm{r}'\right|^3} 
- h(t) \int d\bm{r} \phi(\bm{r}),
\end{eqnarray}
where $H(\phi(\bm{r}))$ is a model of the energy function of the system, random variable $\lambda\left(\bm{r}\right) \sim 1 + \frac{N(0, \textcolor{black}{\nu}^2)}{4}$ represents the spatial heterogeneity of the magnetic material due to defects in the crystal structure or the presence of impurities, and $\textcolor{black}{r_0}$ is the cutoff length set to $\textcolor{black}{r_0}=\frac{2}{\pi}$. 
The strength of the magnetic anisotropy is controlled by $\alpha$, the exchange interaction energy is $\beta$, and the dipole interaction energy is $\gamma$. 
The external magnetic field $h(t)$ is assumed to sweep with time $t$ as follows:
\begin{equation}
h(t)=\begin{cases}
  h_{\rm ini} - vt \:\: (t \leq \frac{h_{\rm ini}}{v})\\
  0.0 \:\: (t > \frac{h_{\rm ini}}{v}).
\end{cases}
\end{equation}
Our simulation was assumed to run until the time at which $t=2T_0 =\frac{2 \times h_{\rm ini}}{v}$, where $T_0 := \frac{h_{\rm ini}}{v}$ is the time at which the external magnetic field $h(t)$ is zero.
Detailed numerical calculation methods are discussed in \ref{appendix_TDGL}. \par
The TDGL equation changes the formation pattern depending on the model parameters $\alpha$, $\textcolor{black}{\nu}$, $\beta$, $\gamma$, and $v$. 
Kudo and Nakamura~\cite{kudo2007field} reports that $\frac{a_1\gamma}{2\beta}$ determines the rough scale of the domain structure, and $\alpha$ controls the variability of the domain size (see \ref{appendix_TDGL_depend}). 
This signifies that even if $\beta$ and $\gamma$ are fixed, we can still generate some variety of patterns by controlling $\alpha$ in the range that can be explained by linear amplification without an external magnetic field. 
From this consideration, Kudo and Nakamura~\cite{kudo2007field} fixed $\frac{a_1\gamma}{2\beta}$, which specifies the scale of the domain structure, to $1$ and focused on the $\alpha$ dependence of the domain structure. 
Kudo and Nakamura also focused on the $v$-dependence controlling the nonequilibrium effects of the domain structure formation dynamics. 
In this study, we also focus on the change in pattern formation dynamics due to $\alpha$ and $v$. 
Specifically, $v$ is set as $0.0001\leq v\leq 0.01$ and $\alpha$ is set as $1.0\leq \alpha \leq 4.0$ at uniformly random or equal intervals depending on the purpose of each analysis. 
Other parameters are set according to Ref.~\cite{kudo2007field} (Table~\ref{tbl1}). 
Note that, in this study, the Hamiltonian parameters $\beta$, $\gamma$, $h_{\rm ini}$, $v$, $\textcolor{black}{r_0}$, and $\textcolor{black}{\nu}$ are fixed, and only $\alpha$ and $v$ are controlled in the ranges described in Table~\ref{tbl1}.\par

\begin{table}[t]
\caption{\label{tbl2}%
Parameters of TDGL equation. 
In this study, the Hamiltonian parameters $\beta$, $\gamma$, $h_{\rm ini}$, $v$, $\textcolor{black}{r_0}$, and $\textcolor{black}{\nu}$ are fixed, and only $\alpha$ and $v$ are controlled in the ranges shown below.}
%\begin{ruledtabular}
\scalebox{0.9}[0.9]{
\begin{tabular}{cccccccc} \hline\hline
$\alpha$&
$\beta$&
$\gamma$&
$h_{\rm ini}$&
$v$&
$\textcolor{black}{r_0}$&
$\textcolor{black}{\nu}$&
Lattice size\\ \hline
{\bf 1.0--4.0} & 2.0 & $\frac{2.0}{\pi}$ & 1.5 & {\bf 0.0001--0.01} & $\frac{\pi}{2}$ & 0.3 & 512 $\times$ 512 (periodic boundary)\\ \hline\hline
\end{tabular}
}
%\end{ruledtabular}
\label{tbl1}
\end{table}

\section{Analysis methods}
\label{sec_method}
\subsection{TDA}
\label{sec_ph}

\begin{figure}[htbp]
  \begin{center}
   \includegraphics[width=\linewidth]{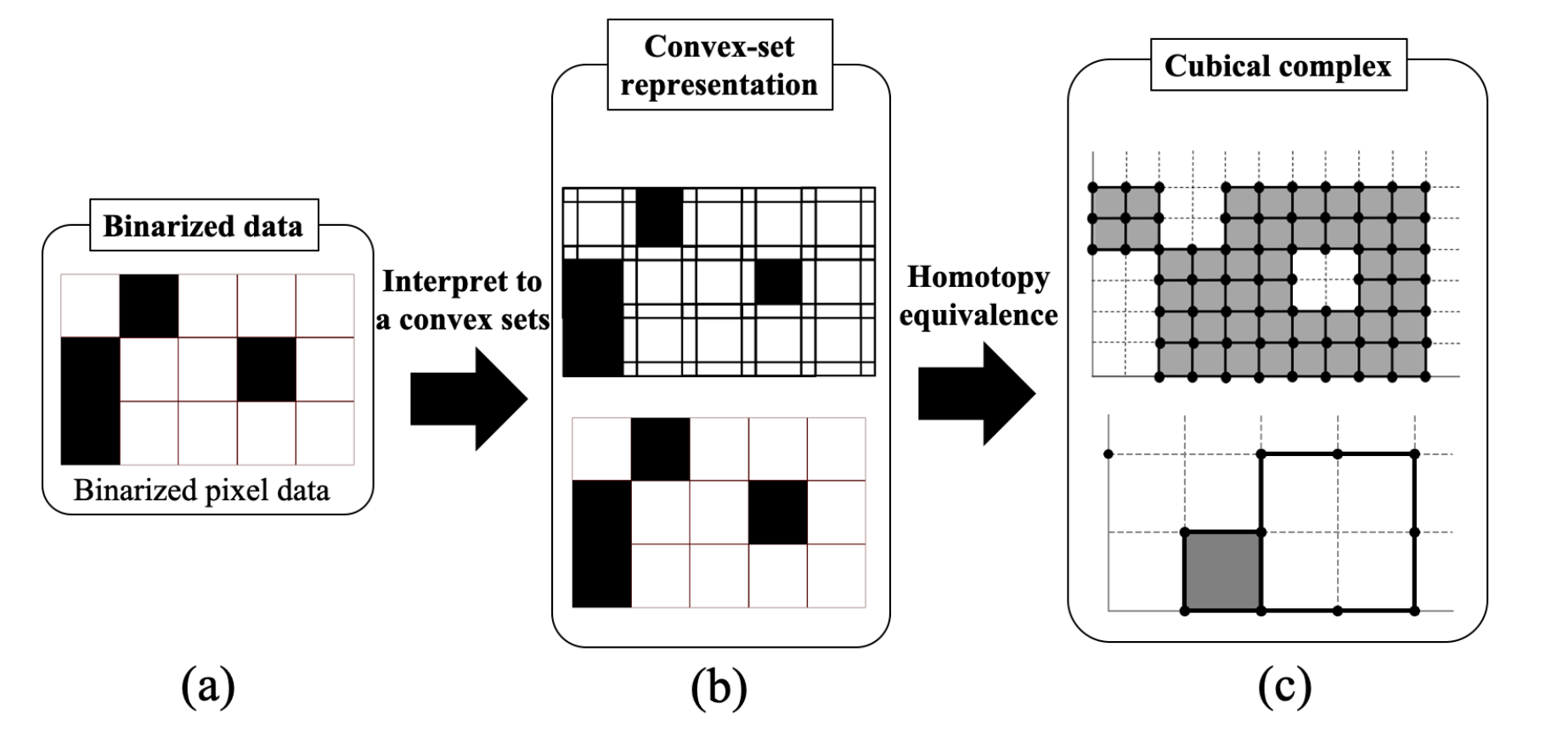}
  \caption{
  Two examples of converting data into a cubical complex (see \ref{appendix}). (a) Binarized pixel image data. \textcolor{black}{The white area is defined as the figure structure.} (b) Two ways of representing the data as closed convex sets (quadrilaterals). \textcolor{black}{%上段は、一つの白いピクセルがある場合に、その中心と周囲に９個の凸閉集合（四角）を割り当てる表現方法で、斜め方向のピクセルを繋がった図形として表現する。
  The upper panel is a representation scheme that assigns nine convex closed sets (squares) to the center and surroundings of a single white pixel, representing the diagonal pixels as connected figures. 
  %下段は、白いピクセルの位置に一つの凸閉集合（四角）を割り当てる表現方法で、ピクセル画像をそのまま凸閉集合と捉える表現方法である。
  The bottom row is a representation scheme that assigns a single convex closed set (square) to the position of a white pixel, which is a way of representing a pixel image itself as a convex closed set as it is.
  } (c) Two ways of representing closed convex sets as a cubical complex. \textcolor{black}{A black dot represents a zero-dimensional cube, a black line represents a one-dimensional cube, and a gray area represents a two-dimensional cube.}}
  \label{fig2}
  \end{center}
\end{figure}

A persistent homology group represents a figure as a set of creation and annihilation of hole structures depending on a single parameter. 
The magnetic domain pattern $\phi\left(\bm{r}\right)$ obtained using the TDGL equation is understood as a two-dimensional grayscale image $G:=\{G(i,j)\in [0,1]\mid\: 1\leq i \leq \textcolor{black}{p}_x, 1\leq j \leq \textcolor{black}{p}_y\}$ composed of $\textcolor{black}{p}_x\times \textcolor{black}{p}_y$ pixels, where $\:G(i,j):\mathbf{N}^2\rightarrow [0,1],\:i=1,2\cdots \textcolor{black}{p}_x,\:j=1,2\cdots \textcolor{black}{p}_y$. 
Thus, if we set the domain as an area that takes a large value, the domain area is given as the set of pixels that are greater than a certain threshold $t$ (superlevel set) as follows:
\begin{equation}
    \{G\textcolor{black}{(i,j)} \mid G(i,j) \geq t, 1\leq i \leq \textcolor{black}{p}_x,1\leq j \leq \textcolor{black}{p}_y,\}.
\end{equation}
By representing one pixel $G(i,j)$ in domain area as a closed convex set $s_{ij}$ [Fig.~\ref{fig2} (b)], we obtain closed convex sets representing domain areas (see \ref{appendix} for more detail).
\begin{equation}
    S_t := \{s_{ij} \mid G(i,j) \geq t, 1\leq i \leq \textcolor{black}{p}_x,1\leq j \leq \textcolor{black}{p}_y,\}
\end{equation}
By reducing the parameter $t$ from 1 to 0, we find that the level set of the grayscale image $S_t$ shows an increasing sequence $\mathbf{S}$ of its geometric elements: 
\begin{eqnarray}
\mathbf{S} : S_{t_1} \subset S_{t_2} \cdots \subset S_{t_{n_{\Theta}}},
\end{eqnarray}
where $1 \geq t_k > t_{k+1} \geq 0$ and $t_{n_{\Theta}}$ is a threshold at which all pixels become larger than it. 
For simplification of notation, the threshold value $t$ is replaced by $t'=1-t$ to let the threshold value correspond to a single parameter that increases corresponding to the increasing sequence of its geometric elements. 
For a two-dimensional grayscale image, zero-dimensional and one-dimensional holes are considered. 
Note that holes in zero dimensions represent a connected component, holes in one dimension represent a ring, and holes in two dimensions represent a cavity. 
When a grayscale image $G$ is binarized into 0 and 1 at a certain threshold $t'$, if 1 is the domain, the 0 region surrounded by 1 is a one-dimensional hole. \par
\begin{figure}[htbp]
  \begin{center}
   \includegraphics[width=\linewidth]{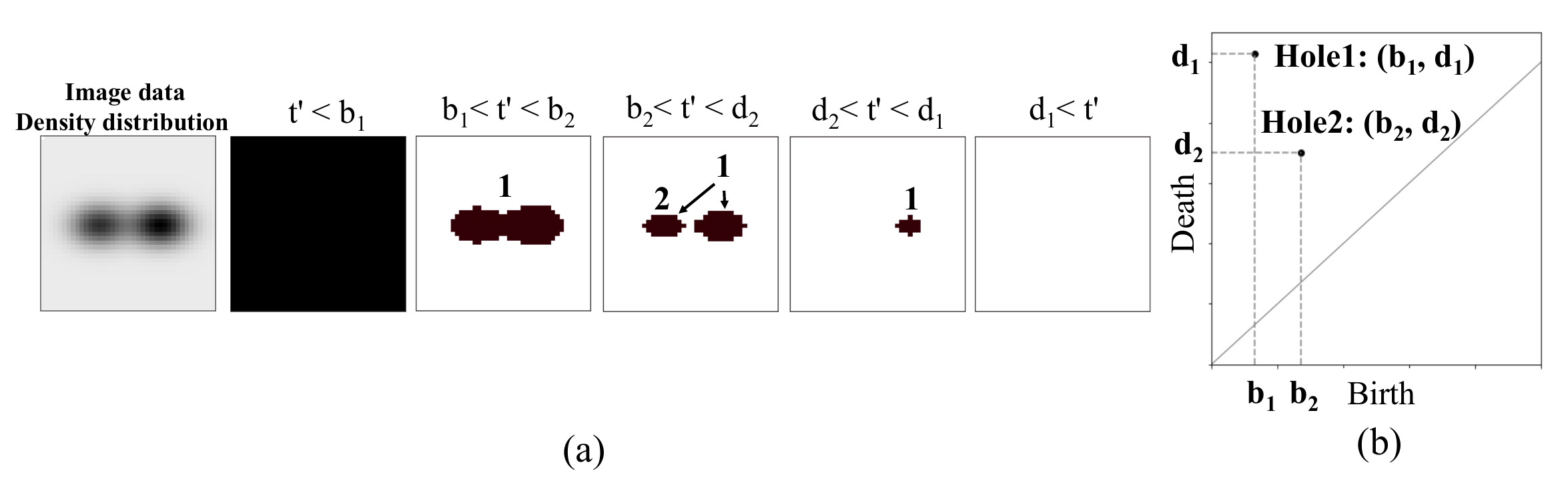}
  \caption{
  Grayscale image data and persistent homology. 
  \textcolor{black}{%この図では、0から1の値を取るピクセル画像をある閾値$t$で２値化した場合のsuperlevel set（白い領域）を図形構造、それに囲まれた領域を穴と定義しています。
  In this figure, a superlevel set (white area) based on a certain threshold $t$ of a pixel image that takes values between 0 and 1 is defined as a figure structure, and the sublevel set (black area) surrounded by the white area is defined as a hole. 
%そして、２値化の閾値$t$を減少（$t'=1-t$を増大）させた際のsuperlevel setの増大列に対するパーシステントホモロジーを考えています。
Then, we consider the persistent homology for an increasing sequence of superlevel sets when the binarization threshold $t$ decreases (increasing $t'=1-t$).}
  (a) One-parameter (threshold) change in the figure structure and (b) the corresponding persistence diagram (PD).}
  \label{fig1}
  \end{center}
\end{figure}
As an example of how the hole structure changes with a single parameter $t'$, we consider a grayscale image illustrated in Fig.~\ref{fig1}(a). 
In this grayscale image, at $t'=0$, the entire pixel image is a nondomain region. 
As $t'$ increases, domain regions emerge and are connected to each other; then, they generate one large hole at $t'=b_1$  [see Hole~1 in Fig.~\ref{fig1}(a)]. 
As $t'$ increases further, the large hole is divided into two holes in a domain at $t'=b_2$  [see Hole~2 in Fig.~\ref{fig1}(a)]. 
When $t'$ is further increased, one of the holes is filled and disappears at $t'=d_2$. 
The other hole is also filled and disappears at $t'=d_1$. 
Note that the death of a hole signifies that the entire hole area at the birth of the hole is filled. 
This implies that even if the hole splits into two owing to an increase in $t'$, the identity is not inherited by one hole, but by both hole regions [see Hole~1 in Fig.~\ref{fig1}(a)]. 
In addition, the remaining hole birth--death structures that cannot be expressed by the birth--death structure of the first hole are treated as another birth--death structure of the second hole [see Hole~2 in Fig.~\ref{fig1}(a)]. 
By defining the birth and death of a hole in this way, we can uniquely determine the set of birth--death pairs of the holes for an increasing sequence of geometric elements of the figure. \par
Persistent homology groups have an isomorphic representation called persistence diagram (PD)~\cite{zomorodian2005computing}. 
PD is represents a set of birth--death pairs with the horizontal axis as the threshold $b_k$ at which an $l$-dimensional hole is generated and the vertical axis as the threshold $d_k$ at which the $l$-dimensional hole disappears as defined below: 
\begin{eqnarray}
\label{PD}
PD_l(\mathbf{K}) = \{(b_k,d_k)\in \mathbf{R}^2\mid k=1,2\cdots N_{\rm hole}\},
\end{eqnarray}
where $\mathbf{K}$ is the increasing sequence of a cubical complex [Fig.~\ref{fig2}(c)] corresponding to the increasing sequence of closed convex sets $\mathbf{S}$ (see \ref{appendix} for more detail). 
The cubical complex is a figure structure formed by pasting together \textcolor{black}{$d$}-dimensional rectangles, such as points, lines, and rectangles. 
The $(b_k,d_k)$ corresponding to one hole on PD is also called a generator of PD, 
which is referred to simply as the ``generator'' in this paper. 
In this study, we perform feature extraction from PD. 
Note that the obtained PD changes depending on how to define the closed convex set $s_{ij}$.
Please refer to \ref{appendix} for information about its definition in this study. 
\textcolor{black}{Note also that the simulation data used in this study were generated under periodic boundary conditions. 
Therefore, the periodic boundary conditions are also taken into account when calculating the PD.}\par
The PDs of $l=0, 1$ are calculated from the magnetic domain pattern, which is understood as two-dimensional grayscale image data. 
In this study, we focus on the PD of $l=1$, which expresses the birth--death pair of a one-dimensional hole (ring) structure. 
Note that, as in the method described for grayscale images, we define a domain as a region that is above a threshold value of magnetic moment. 
To apply machine learning analysis, the obtained $PD_1(\mathbf{K})$ [Eq.~\eqref{PD}], which is the point cloud representing birth--death pairs, is converted into vector data. 
In this study, we vectorize $PD_1(\mathbf{K})$ by a method basd on the persistent image method~\cite{adams2017persistence}. 
In this method, the histogram of the point cloud $PD_1(\mathbf{K})$ is calculated first. 
\begin{eqnarray}
h\left(x,y;PD_1(\mathbf{K}),\sigma\right) = \sum_{i=1}^{N_{\rm hole}} \frac{1}{2\pi\sigma}\exp\left\{-\frac{1}{4\sigma^2}\left[(x - b_i)^2 + (y - d_i)^2\right]\right\}
\end{eqnarray}
We do not correct the histogram to reduce the intensity of short-lived holes corresponding to noisy structures, which is often performed in the persistent image method. 
This is because in this study, we also focus on the information of intradomain fluctuations corresponding to noisy structures. 
The number density is extracted from the histogram function into a $\textcolor{black}{M}\times \textcolor{black}{M}$ grid of $\{x_i,y_j\}_{i,j}$, and the vector data are generated as follows:
\begin{eqnarray}
\label{vector}
{\bf v}(h\left(\:\cdot\:,\:\cdot\:;PD_1(\mathbf{K}),\sigma\right);d) = \left(h(x_1,y_1),h(x_2,y_1),\dots,h(x_d,y_d)\right).
\end{eqnarray}
The kernel width $\sigma$ of the Gaussian kernel used for histogramming and the number of grids $\textcolor{black}{M}$ for vectorization were set as hyperparameters. 
These hyperparameters are set to minimize the validation error of the regression model $F(\:\cdot\:;\boldsymbol{\Theta})$ given by the following equation for the inverse estimation of model parameters described in Sec.~\ref{sec_result1}:
\begin{eqnarray}
E(\boldsymbol{\Theta},\sigma,\textcolor{black}{M}) = \frac{1}{2N_{\rm val}}\sum_{\{{\bf v}_i(\sigma,\textcolor{black}{M}), y_i\}\in D_{\rm val}} \left[F({\bf v}_i(\sigma,\textcolor{black}{M});\boldsymbol{\Theta}) - y_{i}\right]^2,
\end{eqnarray}
where ${\bf v}_i(\sigma,\textcolor{black}{M})$ is the vectorized data of PD for a certain magnetic domain pattern data $i$, $y$ represents the model parameters of the TDGL equation that we want to predict, $D_{\rm val}$ is the validation dataset that is not used for training the regression model, and $\boldsymbol{\Theta}$ is the model parameter of the regression model. 
%The optimized vector dataset are also used for a classification analysis.} 

\subsection{Analysis procedure to reveal the mechanisms of pattern formation dynamics}
\label{sec_anal.proc.}
\begin{figure}
  \begin{center}
   \includegraphics[width=\linewidth]{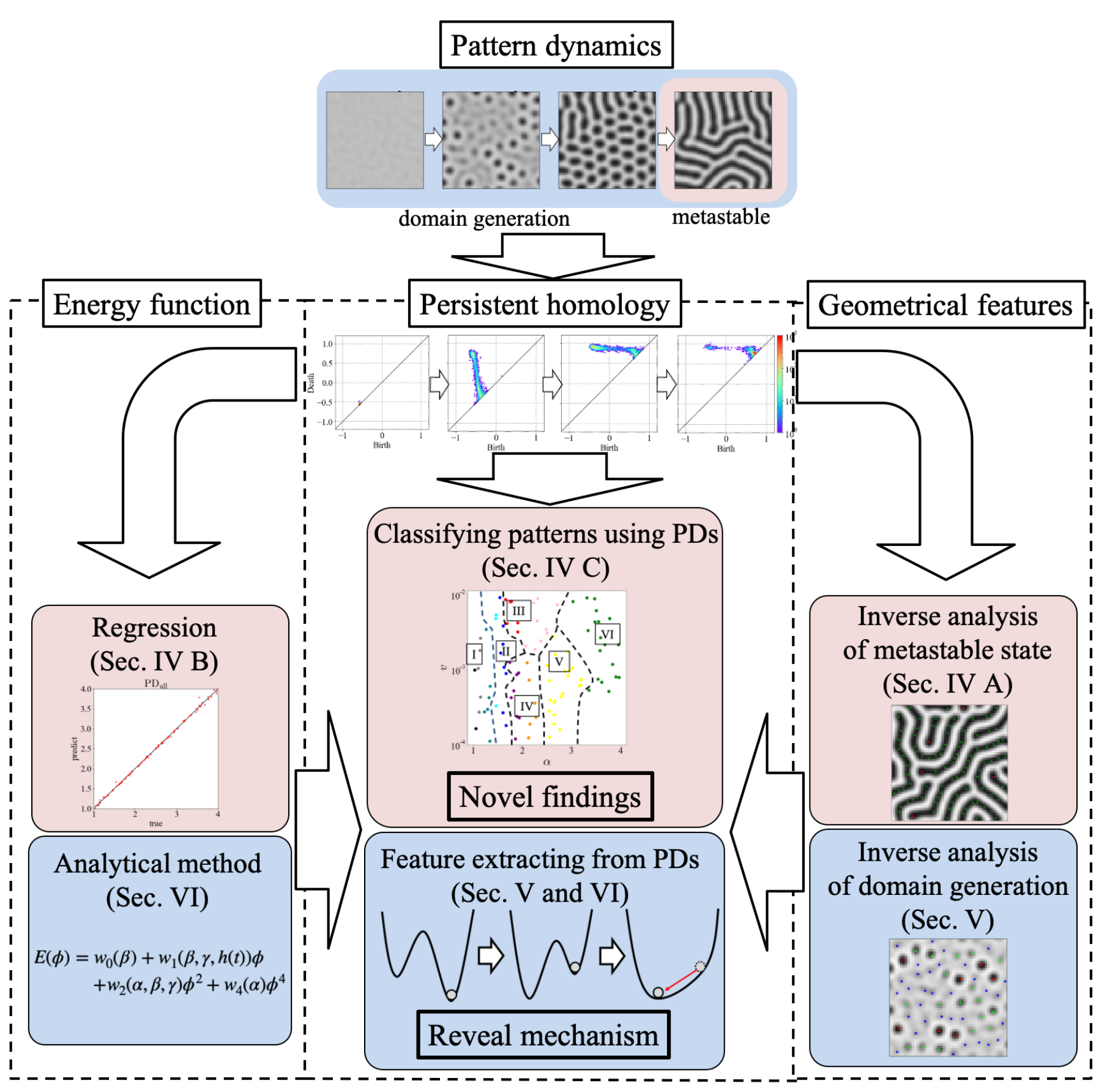}
  \caption{Analysis procedure to reveal the mechanisms of pattern formation dynamics. \textcolor{black}{The analysis procedure described in Sec.~\ref{sec_anal.proc.} and the corresponding sections of this manuscript are summarized in the figure. 
  In the proposed framework, the analysis is performed to map the geometric structure of the pattern dynamics to the energy function that represents the mechanism of the system according to the arrows through TDA.}}
  \label{fig_framework}
  \end{center}
\end{figure}
To understand the mechanism of the process of magnetic domain pattern formation, it is appropriate to analyze a snapshot taken at a characteristic time representative of the system. 
Pattern formation processes with long-range interactions have a complex multivalley energy landscape. 
Therefore, depending on the initial state of the system or the interaction process of the system with the environment, the system should reach different metastable states or saddle points. 
In this study, the initial state of the system was set to be the same in all simulations. 
The system is expected to reach different local minima or saddle points depending on the sweep rate $v$ of the external magnetic field. 
Therefore, the snapshot at $t=2T_0$, which is the time when the system behaves with a certain degree of stability after the sweep of the external magnetic field ends, was chosen for analysis. 
By analyzing the snapshot at $t=2T_0$, we verify that the PD retains sufficient information about the control parameters of the system, and we perform classification of magnetic domain patterns based on topological features and machine learning methods to elucidate the relationship between geometric features and control parameters.\par
The physical mechanism underlying the classification of magnetic domain patterns obtained from the analysis of the snapshot at $t=2T_0$ could be understood by analyzing the formation process of the each pattern class. 
For this purpose, we calculated the PD corresponding to each snapshot at each time of the pattern formation process and searched for the time transition of the geometric structure associated with the classification result obtained by analyzing the snapshot at $t=2T_0$. 
By investigating the correspondence between the obtained time transition properties of the geometric structure and the energy model of the TDGL equation, we discover the interpretable mechanism underlying the pattern formation of magnetic domain.\par
The analysis procedures described above and the corresponding sections are summarized in Fig~\ref{fig_framework}.

\section{Analysis of snapshot at $t=2T_0$}
\label{sec_result1}
%From the results of TDA of the stable-state magnetic domain pattern, we identified the geometrical structure that characterizes the control parameters} $\alpha$ and $v$ of the TDGL equation, and we search for novel insights into the relationship between the geometric features of the magnetic domain and the physical properties of the system. 
In this section, we first use the inverse analysis method of PD~\cite{obayashi2018persistence, vandaele2020topological}, which we will explain later, to obtain a geometrical interpretation of the characteristic structure of the PD of the magnetic domain pattern. 
Next, we conduct regression analysis to verify whether the PD retains sufficient information about the control parameters $\alpha$ and $v$, which are the dominant parameters of our domain formation system. 
Furthermore, by utilizing clustering methods to create a pattern classification diagram in the $\alpha$--$v$ plane, we interpret the geometrical features extracted from PD. 
In addition to the verification, we performed a classification of magnetic domain structures based on topological features.
The results of these analyses are described below.\par

\subsection{PD and corresponding geometrical features}
\label{result_stable}
\begin{figure}[htbp]
  \begin{center}
   \includegraphics[width=\linewidth]{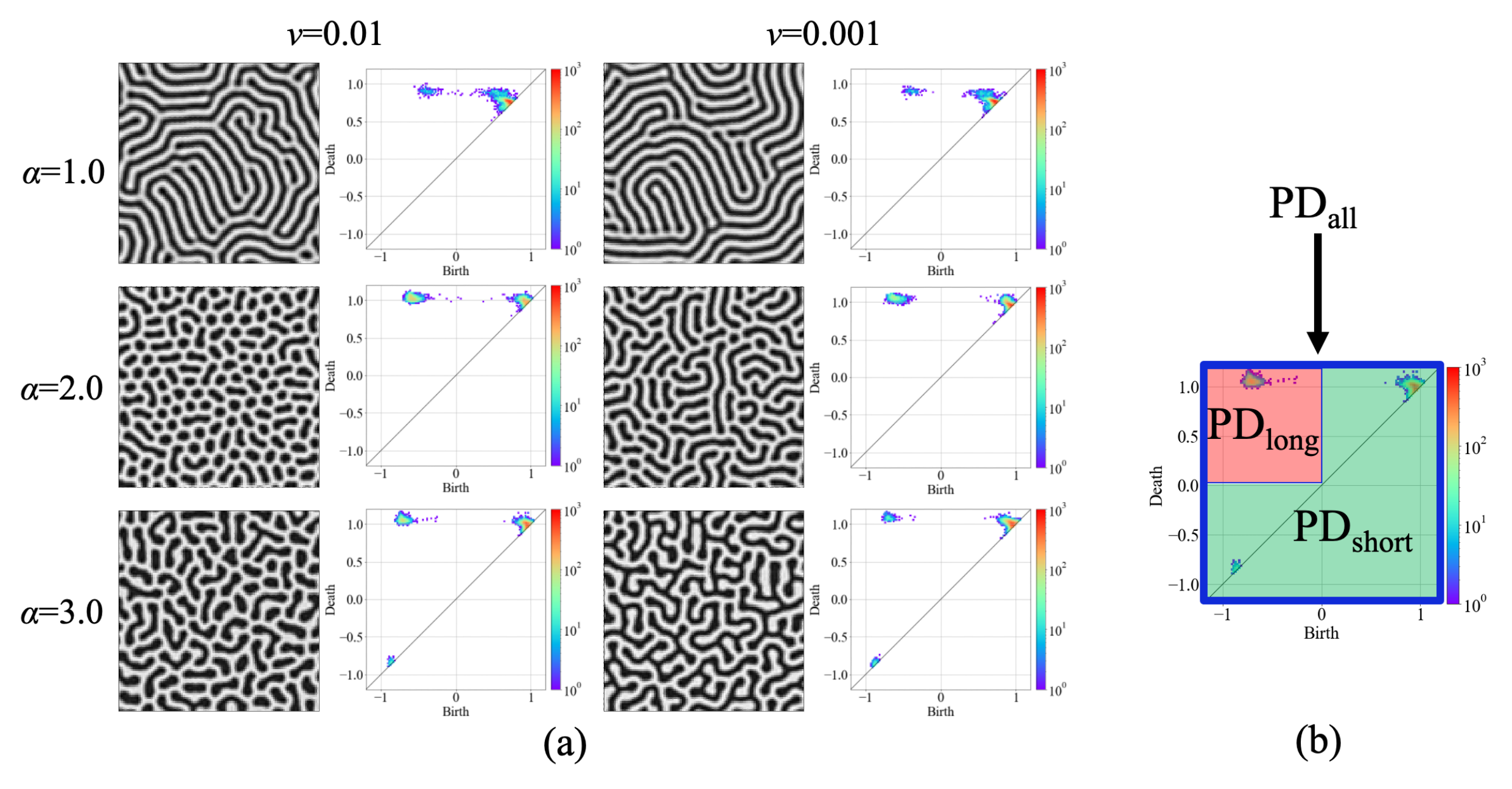}
  \caption{(a) Magnetic domain patterns (first and third columns) and corresponding PDs (second and fourth columns) for the magnetic domain patterns at $t=2T_0$ for each of $\alpha$ and $v$.
  (b) Definitions of regions of PD: ${\rm PD_{all}}$, ${\rm PD_{long}}$, and ${\rm PD_{short}}$.}
  \label{fig3}
  \end{center}
\end{figure}
We performed numerical simulations based on the TDGL equation for $\alpha = \{1.0,2.0,3.0\}$ and $v=\{0.01,0.001\}$, and we obtained the magnetic domain patterns as shown in the first and third columns of Fig.~\ref{fig3}(a) at $t=2T_0$. 
%\textcolor{black}{ここで改めて言及しておくが、magnetic domain pattern図は、実数値を取る磁気モーメント$\phi\left(\bm{r}\right)$を、最大値を白、最小値を黒としたグレースケール画像として表現したものである。}
\textcolor{black}{It is worth mentioning here again that the magnetic domain pattern figure is a grayscale image of the magnetic moment $\phi\left(\bm{r}\right)$ that takes real values, with the maximum value in white and the minimum value in black.} 
The PDs corresponding to each magnetic domain pattern are shown in the second and fourth columns in Fig.~\ref{fig3}(a). 
%\textcolor{black}{PDは2値化の閾値を$t$とした場合に、$t'=-t$の増大に応じた$\phi\left(\bm{r}\right)$のsuperlevel-setの増大列に対する1次のPHとして算出された。
%これは、$-\phi\left(\bm{r}\right)$のsublevel-setの増大列に対する1次のPHであることに他ならない。
%ちなみに、先行研究との対応やわかりやすさから、$\phi\left(\bm{r}\right)$を、最大値を白、最小値を黒としたグレースケール画像として描画することから、前述の閾値$t$を$t'$に置き換えるような回りくどい表現となっていることに注意されたい。}
\textcolor{black}{In this study, we calculated the PD corresponding to the one-dimensional PH for the increasing sequence of superlevel sets of the magnetic moment distribution $\phi\left(\bm{r}\right)$ as $t'=-t$ increases, where $t$ denotes the binarization threshold of $\phi\left(\bm{r}\right)$ that is the grayscale image.
This is simply a one-dimensional PH for the increasing sequence of sublevel sets of $-\phi\left(\bm{r}\right)$. 
Note that we used a roundabout expression such as replacing the previously mentioned threshold $t$ with $t'$ to draw $\phi\left(\bm{r}\right)$ as a grayscale image with the maximum value as white and the minimum value as black for the sake of consistency with previous studies and clarity~\cite{jagla2004numerical, kudo2007magnetic}.}
From this figure, we can see that PDs tend to have a different distribution structure depending on $\alpha$ and $v$. 
It is also observed in this figure that the generators are concentrated \textcolor{black}{on more than two of the three regions} of PD: lower left, upper left, and upper right, regardless of $\alpha$ and $v$. 
In general, the generator around the diagonal line of PD represents a short-life hole, i.e. the birth and death thresholds are close, whereas the generators far from the diagonal line represent long-life holes, i.e., the birth and death thresholds are far apart. 
We investigated the geometrical structure of the magnetic domain pattern corresponding to the long-life region ${\rm PD_{long}}$ and the short-life region ${\rm PD_{short}}$ [Fig.~\ref{fig3}(b)]. 
\textcolor{black}{%長寿命・短寿命領域を寿命で定義するのではなく、${\rm PD_{long}}$と${\rm PD_{short}}$と定義するのは、$\alpha$によってgeneratorの分布が変動するため、寿命による定義では、長寿命のgenerator（左上）と短寿命のgenerator（右上、左下）を$\alpha$によらず統一的に抽出する基準を設定できないためです。
The long-life and short-life regions are respectively defined as ${\rm PD_{long}}$ and ${\rm PD_{short}}$ instead of by lifetime, because the distributions of generators vary depending on $\alpha$, and the definition by lifetime does not allow us to set the criteria for extracting the distributions of long-life and short-life generators in a unified manner regardless of $\alpha$. 
}\par
\begin{figure}[htbp]
  \begin{center}
   \includegraphics[width=0.9\linewidth]{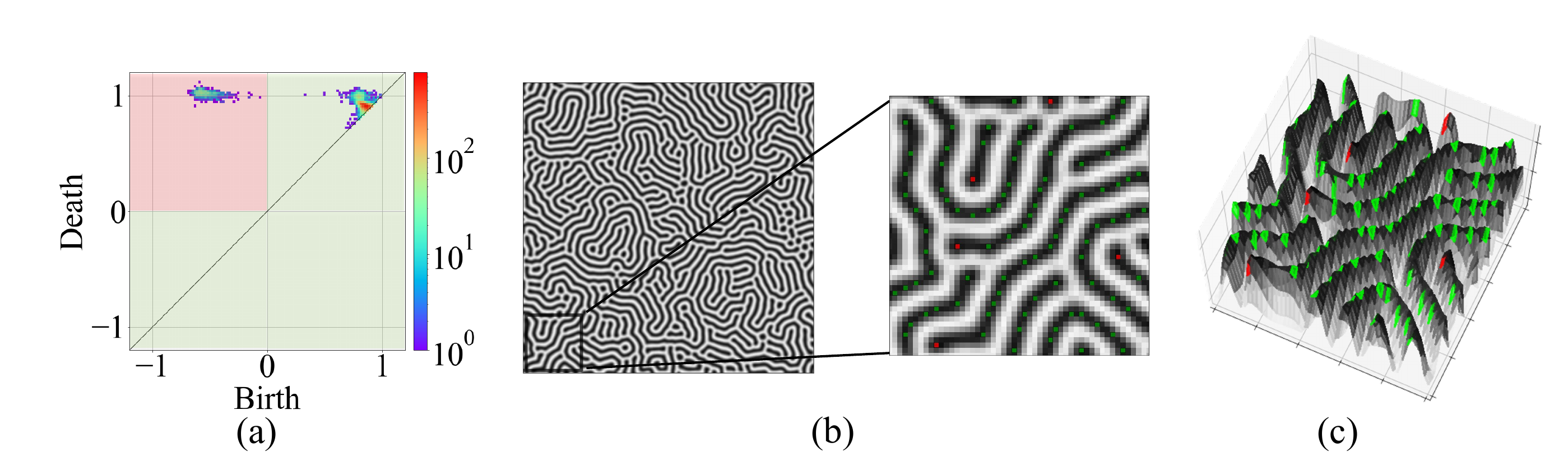}
  \caption{Correspondence between the distribution structure of the generators in PD and the magnetic domain pattern. (a) PD of the magnetic domain pattern with $\alpha =1.5$, $v=0.01$ at $t=2T_0$. 
 The long-life region of PD is shown in red, and the short-life region of PD is shown in green. 
  (b) Magnetic domain pattern and its enlarged image. 
  Red dots in the enlarged image represent the positions of the death cubes corresponding to the long-life region of PD, whereas green dots represent the positions of death cubes corresponding to the short-life region of PD. 
  (c) Three-dimensional representation of the \textcolor{black}{inverse magnetic domain pattern $-\phi\left(\bm{r}\right)$}, where red and green dots represent the locations of long-life and short-life death cubes, respectively. 
  \textcolor{black}{
  %このように、赤点は独立した連峰にそれぞれ一つ存在する。
  Thus, there is one red point in each of the independent domains. 
  %つまり、赤点の数は$\phi\left(\bm{r}\right)=0$で２値化した場合のBetti数に対応する。
  That is, the number of red dots corresponds to the Betti number binarized by an appropriate value such as $\phi\left(\bm{r}\right)=0$.}
  The HomCloud library~\cite{homcloud} was used to extract the positions of the death cubes.
  The method to construct convex closed sets to estimate the positions of the death cubes using the HomCloud library is described in \ref{appendix}.}
  \label{fig4}
  \end{center}
\end{figure}

We used the inverse analysis method~\cite{obayashi2018persistence, vandaele2020topological} to geometrically interpret the characteristic structure of generator distribution in the PD of the magnetic domain pattern. 
In particular, we extracted the last annihilating cube in one cubical complex, which disappears when the threshold value changes, as the approximate location of the hole corresponding to a certain generator. 
Hereinafter, we call the cube a ``death cube''. 
As a result, it is found that the death cubes of the long-life region ${\rm PD_{long}}$ are the pixels with the smallest magnetic moment in the same magnetic domain, and the death cubes of the short-life region ${\rm PD_{short}}$ are the pixels at a relatively small peak existing in the magnetic domain. 
\textcolor{black}{In other words, ${\rm PD_{long}}$ and ${\rm PD_{short}}$ correspond to high-contrast and low-contrast structures of a grayscale image $\phi\left(\bm{r}\right)$, respectively.}
It means that the long-life region ${\rm PD_{long}}$ represents the magnetic domain structure, and the short-life region ${\rm PD_{short}}$ represents fluctuations in the inverse \textcolor{black}{magnetic} domain of $\phi < 0$ (Fig.~\ref{fig4}).
A similar inverse analysis shows that the short-life generators distributed in the lower left of the PD, as seen in Fig.~\ref{fig3} at $\alpha=3.0$, correspond to fluctuations in the positive magnetic domain of $\phi > 0$.\par

\subsection{Regression analysis of PDs}
\label{sec_regress}
\begin{figure}[htbp]
  \begin{center}
   \includegraphics[width=\linewidth]{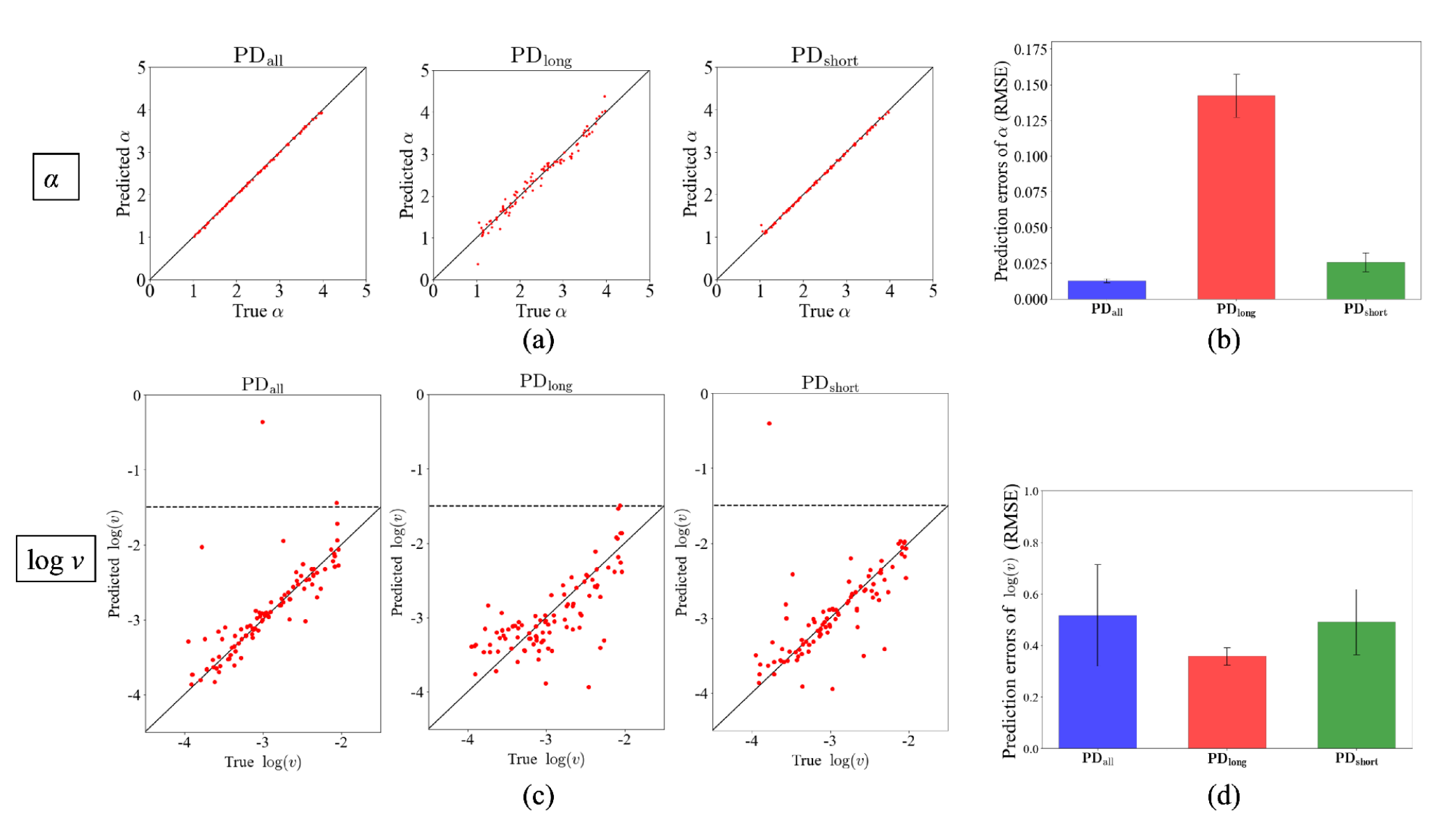}
  \caption{(a) Regression results for $\alpha$. The horizontal axis is the true value of $\alpha$ used for the TDGL simulation, whereas the vertical axis is the prediction value estimated in the outer loop of nested cross-validation. (b) Prediction error (root mean squared error) of the constructed regression model for $\alpha$ when Ridge regression is applied with vectorized ${\rm PD}_{\rm all}$, ${\rm PD}_{\rm long}$, and ${\rm PD}_{\rm short}$ as inputs. (c) Regression results for $\log\:v$. The horizontal axis is the true value of $\alpha$ used for the TDGL simulation, whereas the vertical axis is the prediction value estimated in the outer loop of nested cross-validation. (d) Prediction error (root mean squared error) of the constructed regression model for $\log\:v$ when Ridge regression is applied with vectorized ${\rm PD}_{\rm all}$, ${\rm PD}_{\rm long}$, and ${\rm PD}_{\rm short}$ as inputs. }
  \label{fig5}
  \end{center}
\end{figure}

We conducted regression analysis to verify whether PD retains sufficient information to estimate the control parameters $\alpha$ and $v$ of the system. 
For regression analysis, $\alpha$ and $\log v$ were generated \textcolor{black}{randomly from a uniform distribution} in the ranges of $\alpha=[1.0,5.0]$ and $\log\:v=[-4,-2]$, and the TDGL equation was run numerically using these parameters until $t=2T_0$. 
96 sets of simulation data $\{G_i,\alpha_i,v_i\}_{i=1}^{96}$ were analyzed, where the numerical calculation errors did not diverge owing to rapid changes in the magnetic moments $\phi\left(\bm{r}\right)$ during the formation process. 
We apply multiple regression analysis to the data set $\{{\bf v}_i(\sigma,\textcolor{black}{M}), \alpha_i\}_{i=1}^{96}$ and $\{{\bf v}_i(\sigma,\textcolor{black}{M}), \log\:v_i\}_{i=1}^{96}$, where ${\bf v}_i(\sigma,\textcolor{black}{M})$ is vectorized PDs generated by the procedure in Sec.~\ref{sec_ph}. 
For the multiple regression analysis, we performed Ridge regression, its hyperparameters and generalization error were estimated by nested cross-validation framework~\cite{cawley2010over} (detailed procedure is described in \ref{sec_nested}). 
${\rm PD}_{\rm all}$ in Fig.~\ref{fig5}(a) shows the $\alpha$ estimation and ${\rm PD}_{\rm all}$ in Fig.~\ref{fig5}(c) shows the $\log\:v$ estimation. 
This regression result confirms that a high prediction accuracy of $\alpha$ [${\rm PD}_{\rm all}$ in Fig.~\ref{fig5}(a)] and the accuracy of $\log\:v$ to some extent [${\rm PD}_{\rm all}$ in Fig.~\ref{fig5}(c)] can be obtained. 
When we employed all the regions of PD, ${\rm PD}_{\rm all}$, the root mean squared error (RMSE) of $\alpha$ was estimated to be about 0.001 [${\rm PD}_{\rm all}$ in Fig.~\ref{fig5}(b)], and its R-squared  value was greater than 0.99. 
Since the range for $\alpha$ is 1--4, this RMSE is very small.  
\textcolor{black}{Note that such high estimation accuracy is not particularly surprising. For example, although, unlike in our study where the TDGL model is simple and does not consider external magnetic fields, it has been reported that similar high estimation accuracy of a parameter by combining feature extraction using TDA and a regression model, as described in Introduction~\cite{calcina2021parameter}.}
The estimation accuracy of $\log\:v$ was not as high as that of $\alpha$. 
We statistically determined whether the linear regression can extract geometric information from the PD, which would allow us to estimate $\log\:v$. 
By performing the permutation test, we confirmed that the accuracy of the estimation of $\log\:v$ is significantly higher than that of the null model (\textcolor{black}{see} \ref{appendix3}). 
As described in Introduction, it has been reported that a high $\alpha$ regression performance can be obtained using TDA in a simple Ginzburg--Landau equation whose energy model does not have a time-varying external magnetic field~\cite{calcina2021parameter}. 
The facts that a similarly high performance was obtained for the system of energy functions with a time-dependent external magnetic field targeted in this study and that the parameter $v$ governing its time-varying term was predicted to some extent support the effectiveness of TDA for the analysis of complex nonequilibrium pattern formation dynamics.\par
Here, we try to elucidate the correspondence of the control parameters $\alpha$ and $v$ to the geometrical structure of magnetic domains obtained from the inverse analysis of PD. 
For this purpose, a regression model predicting $\alpha$ and $\log\:v$ with the overall regions ${\rm PD_{all}}$, the long-life region ${\rm PD_{long}}$, and the short-life region ${\rm PD_{short}}$ as explanatory variables was constructed using the procedure described above. 
As a result, in the estimation of $\alpha$, ${\rm PD_{short}}$ achieved a high prediction performance similarly to ${\rm PD_{all}}$, whereas ${\rm PD_{long}}$ has a relatively low estimation accuracy [Fig.~\ref{fig5}(b)]. 
In the estimation of $\log\:v$, we did not see a clear difference between characteristics as seen in $\alpha$ [Fig.~\ref{fig5}(d)]. 
These results suggest that not only the geometric features of the binarized magnetic domains, as used in previous research, but also the contour features of the magnetic domains, such as fluctuations in the domain, contain information about the physical mechanisms of the system. 
On the other hand, in the estimation of $\log\:v$, the RMSE of ${\rm PD}_{\rm long}$ was not larger than those of other regions, which indicates that ${\rm PD}_{\rm long}$ retains useful information for the estimation of $v$. \par
In our simulation setting, these results suggest that $\alpha$, which dominates the anisotropic energy, contributes to the formation of the local structure of magnetic domains\textcolor{black}{, such as the fluctuations of magnetic moments in the domain}, and $v$, which dominates the nonequilibrium process, contributes to the formation of the global structure of magnetic domains \textcolor{black}{such as the domain pattern binarized  with zero magnetic moments}. 
These are consistent with a previous study by Kudo and coworkers in which $\alpha$ dominates the variation in the periodic structure of the spatial magnetic distribution~\cite{kudo2007magnetic} whereas $v$ dominates the domain size~\cite{kudo2007field}. 
These results can be \textcolor{black}{achieved} because \textcolor{black}{of information such as ${\rm PD}_{\rm short}$ that could not be obtained from the binarized magnetic domain image: in other words, it is important that the magnetic moment $\phi\left(\bm{r}\right)$ is treated as a continuous variable}.\par

\subsection{Classifying patterns based on PDs}
Magnetic domain patterns were classified on the basis of PD, and the relationship between the characteristics representing each classification class and the control parameters $\alpha$ and $v$ of the system was examined. 
We applied the K-means clustering method~\cite{rokach2005clustering, forgy1965cluster, macqueen1967some} to the vectorized data $\{{\bf v}_i(\sigma_{\rm opt},\textcolor{black}{M}_{\rm opt})\}_{i=1}^{96}$ of PDs to find pattern classes with similar properties. 
\textcolor{black}{The Ridge regression employed in the previous section is known to perform weighted regression for each direction of the principal components of a principal component analysis according to the magnitude of its variance~\cite{hastie_09_elements-of.statistical-learning}. 
The K-means clustering method is also known to be affected by principal components with large variances~\cite{ding04}. 
Therefore, the K-means clustering method is appropriate for investigating the mechanism behind the information on PD identified in the regression analysis by Ridge regression.}\par
The vectorized parameters $\sigma_{\rm opt}$ and $\textcolor{black}{M}_{\rm opt}$ of PDs were selected on the basis of regression analysis results in the \textcolor{black}{ Sec.~\ref{sec_regress}}. 
This is based on the idea that a hyperparameter that effectively predicts physical properties that are important for a phenomenon can effectively extract information about the target. 
The classification is based on the Euclidean distance defined for any two vectors ${\bf v}_{i_1}(\sigma_{\rm opt},\textcolor{black}{M}_{\rm opt})$ and ${\bf v}_{i_2}(\sigma_{\rm opt},\textcolor{black}{M}_{\rm opt})$ taken from the vectorized data $\{{\bf v}_i(\sigma_{\rm opt},\textcolor{black}{M}_{\rm opt})\}_{i=1}^{96}$:
\begin{equation}
\sqrt{\left[{\bf v}_{i_1}(\sigma_{\rm opt},\textcolor{black}{M}_{\rm opt}) - {\bf v}_{i_2}(\sigma_{\rm opt},\textcolor{black}{M}_{\rm opt})\right]^T\left[{\bf v}_{i_1}(\sigma_{\rm opt},\textcolor{black}{M}_{\rm opt}) - {\bf v}_{i_2}(\sigma_{\rm opt},\textcolor{black}{M}_{\rm opt})\right]}.
\end{equation}  
The number of clusters $K$ was determined using the elbow method~\cite{thorndike1953belongs}, and $K=11$ was chosen. \par
\begin{figure}[htbp]
  \begin{center}
   \includegraphics[width=0.68\linewidth]{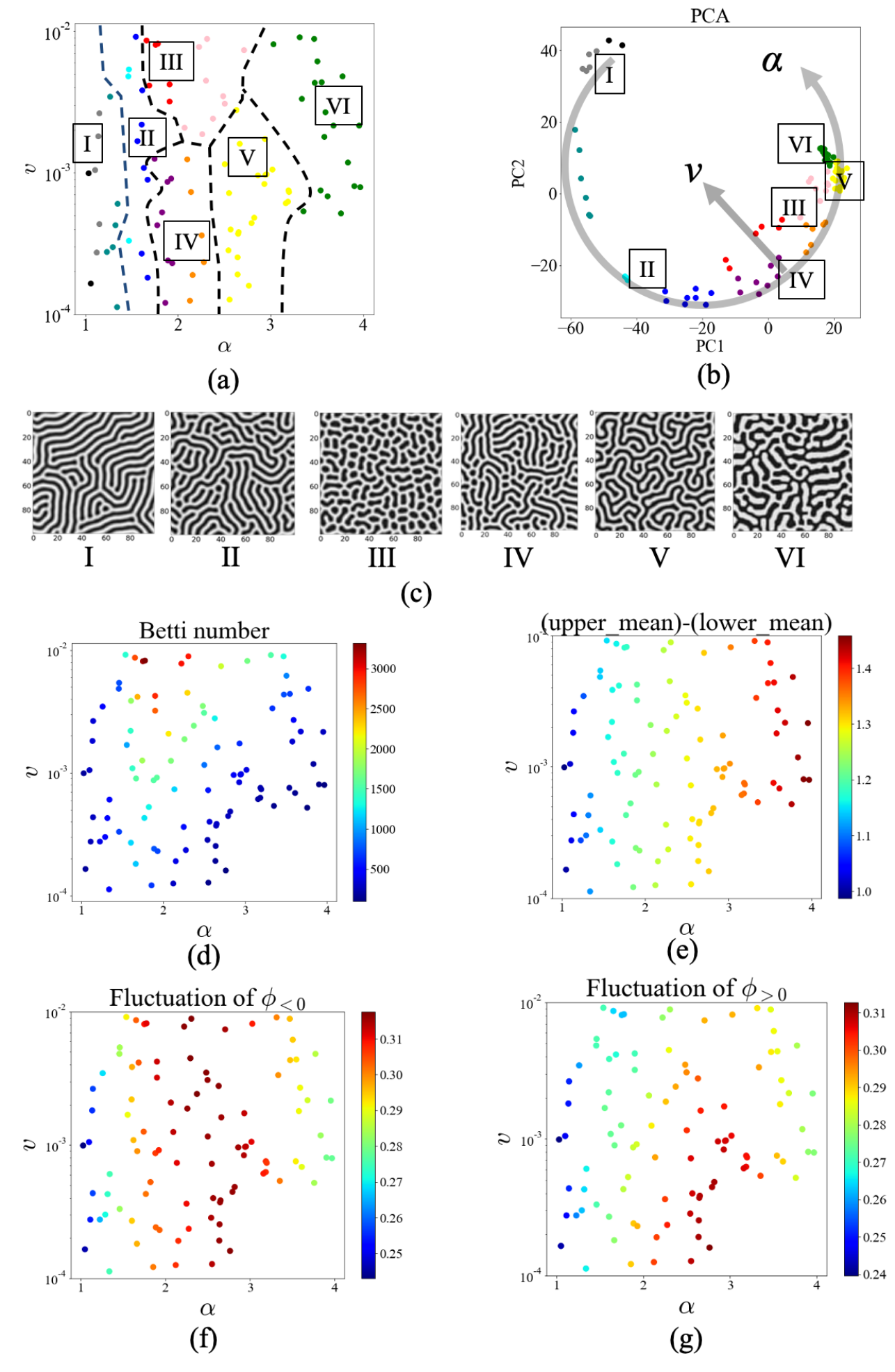}
  \caption{Results of classification of the magnetic domain patterns using feature data $\{{\bf v}_i(\sigma,\textcolor{black}{M})\}_{i=1}^{96}$ [Eq.~\ref{vector}] extracted from PD. (a) Classification results were obtained by K-means clustering. The horizontal axis represents $\alpha$ and the vertical axis represents $v$ of the TDGL equation. (b) Results of PCA of $\{{\bf v}_i(\sigma,\textcolor{black}{M})\}_{i=1}^{96}$. The horizontal axis represents the principal component with the largest eigenvalue, whereas the vertical axis represents the principal component with the second largest eigenvalue. (c) Magnetic domain patterns at $t=2T_0$ corresponding to points \textcolor{black}{around} I--VI in (a) and (b). (d) Distribution of the Betti number of the domain structure binarized using $\phi=0$ at $t=2T_0$. (e) Distribution of differentiation between the mean magnetic moments of the positive magnetic domain and negative magnetic domain at $t=2T_0$. (f) Distribution of the standard deviation of the magnetic moment of the inverse magnetic domain at $t=2T_0$. (g) Distribution of the standard deviation of the magnetic moment of the positive magnetic domain at $t=2T_0$.}
  \label{fig7}
  \end{center}
\end{figure}

The classification result is shown in Fig.~\ref{fig7}(a). 
%Figure~\ref{fig7}(a) takes $\alpha$ as the horizontal axis and $v$ as the vertical axis, and each point in the figure corresponds to the magnetic domain pattern at $t=2T_0$ obtained by numerical simulation using each parameter. 
%The different colors of points in Fig.~\ref{fig7}(a) represent the different clusters. 
From Fig.~\ref{fig7}(a), it was found that the clusters obtained on the basis of PDs also have a cluster structure in the control parameters $\alpha$--$v$ space. 
This supports the results of regression analysis that the geometric structure obtained from the PDs is strongly related to $\alpha$ and $v$. 
These classification results could be independent of the \textcolor{black}{distance-based} clustering method (see \ref{appendix3}). 
Figure~\ref{fig7}(b) shows the distribution of the vectorized PD data $\{{\bf v}_i(\sigma_{\rm opt},\textcolor{black}{M}_{\rm opt})\}_{i=1}^{96}$ reduced to two dimensions by principal component analysis (PCA). 
The cumulative contribution of the two principal components comprising the reduced space was more than 85\%. 
This confirms that the information of $\alpha$ and $v$ is smoothly embedded in the principal component space of the features. 
The increase in $\alpha$ appears to correspond to the \textcolor{black}{counterclockwise} direction in this reduced space, and the increase in $v$ appears to correspond to the direction of the center of the circle. 
Thus, the feature space obtained from the PD appears to retain significant information about the physical property of the system.\par
The obtained clusters are understood physically by defining some feature values derived from discussion of regression analysis results. 
For later explanation, we define six parameter regions delineated by dotted lines in Fig.~\ref{fig7}(a). 
Examples of magnetic domain patterns I--VI for each parameter region are illustrated in Fig.~\ref{fig7}(c). 
Pattern III has a sea-island structure~\cite{kudo2007field, kudo2007magnetic}, and patterns V and VI have a labyrinthine structure~\cite{kudo2007field, kudo2007magnetic}. 
Pattern I also has a labyrinthine structure, and patterns II and IV have a mixture of sea-island and labyrinthine structures. 
Because ${\rm PD_{long}}$ and ${\rm PD_{short}}$ were found to have information on the control parameters $\alpha$ and $v$ of the system from the regression analysis, it should be possible to understand the physical meaning of the cluster structure on the basis of the features that are considered to be related to these PD regions. 
First, we interpret the cluster structure using as a feature \textcolor{black}{the number of domains (Betti number) of the isolated} magnetic domain pattern obtained when binarizing with $\phi = 0$ as the threshold value. 
The Betti number can be calculated from the sum of generators whose birth--death intervals include $t'=0$, which corresponds to the sum of generators in ${\rm PD_{long}}$. 
%The Betti number in the domain pattern binarized at $t'=0$ provides information about the global domain structure, such as the number of islands in a sea or the number of walls in a labyrinthine structure. 
The distribution of the Betti number indicates that regions where pattern III exists can be characterized as regions with a large Betti number [Fig.~\ref{fig7}(d)]. 
This suggests that this region, similar to pattern III, shares a common sea-island structure with a greater number of isolated domains than the labyrinthine structure. 
Next, we employed the difference between the average magnetic moment of the positive magnetic domain (upper mean) and the average magnetic moment of the inverse magnetic domain (lower mean) as the feature to interpret the cluster structure. 
The birth and death times of the generator in the upper right of ${\rm PD_{short}}$ give the range of magnetic moments taken by the inverse magnetic domain, and the birth and death times of the generator in the lower left of ${\rm PD_{short}}$ give the range of magnetic moments taken by the positive magnetic domain. 
From this, the approximate average values of the inverse and positive magnetic domains could be determined. 
The distribution of the difference between the magnetic moments of the positive and negative magnetic domains characterizes the region where pattern V belongs as a region with particularly large values [Fig.~\ref{fig7}(e)]. 
Moreover, the difference between the magnetic moments of the positive and negative magnetic domains increases monotonically with increasing anisotropy parameter $\alpha$ [Fig.~\ref{fig7}(e)]. 
This tendency is also observed in Fig.~\ref{fig3}(a). 
That is, the increase in the anisotropy parameter $\alpha$ causes systematic increase  the birth and death times of the generator in the upper right region of ${\rm PD_{short}}$ [Fig.~\ref{fig3}(a)]. 
This simple one-to-one correspondence between $\alpha$ and simple features on PD may be the reason for the high accuracies of estimating $\alpha$ using only ${\rm PD_{short}}$, which does not have information of the global domain structure. 
Note that the birth and death times of the generator of ${\rm PD_{long}}$ also provide limited information about the magnetic moments of the positive and inverse magnetic domains, that is, \textcolor{black}{the rough magnetic moment value of the positive domain that is expressed as the threshold at which a hole is formed for the first time and the lower limit of the inverse domain for the time when a hole disappears.} 
This property might explain the relatively low accuracy of estimating $\alpha$ using ${\rm PD_{long}}$. 
The last two features are the variances of the magnetic moments within the inverse and positive magnetic domains. 
We define an inverse magnetic domain as a region where $\phi <0$ and a positive magnetic domain as a region where $\phi >0$. 
The number of generators in the upper-right region of PD and the birth and death times of the generators are associated with the variances of the structure within the inverse magnetic domain, and the generator distribution in the lower-left region is associated with the variances within the positive magnetic domain. 
The distribution of the variance of the inverse magnetic domains characterizes the regions containing patterns III--V as those with significantly larger values [Fig.~\ref{fig7}(f)]. 
The distribution of the variance of the positive magnetic domains shows that only the region containing pattern V can be characterized as a region with significantly large values [Fig.~\ref{fig7}(g)]. 
In summary, the parameter region containing pattern III is characterized by the Betti number, the parameter region containing pattern VI by the difference between the magnetic moments of the positive and inverse magnetic domains, and the parameter region containing pattern V by the variance of the magnetic moments of the positive magnetic domains. 
The parameter region containing pattern IV is characterized by a combination of the Betti number and variances of the magnetic moment. 
The parameter region containing pattern I is characterized by particularly small values of three features of the magnetic moment. 
The parameter region containing pattern II is characterized by the absence of a specific value of all features. 
Thus, part of the physical mechanism underlying the \textcolor{black}{result of clustering analysis} on the basis of PD has been elucidated. 
\textcolor{black}{
%これらの分析結果は本研究で得られた結果は、一度現象を記述する既存の特徴量が判明すれば、敢えてPDを元にした特徴量を用いる必要はないことも示唆している。
These analyses also suggest that for the results obtained in this study, PD-based features need not be used once the classical features describing the phenomenon are identified. 
%TDAを用いるbenefitは、PDが多様な情報を含んでおりかつ逆解析による解釈性があるため、TDAの結果からパターンを記述する適切な統計量やフーリエ特徴量のような既存の特徴量へのマップを発見できる点だと考えます。
The benefit of using TDA is that the PD contains a variety of information and is interpretable by inverse analysis, enabling the discovery of maps from TDA results to appropriate classical features that describe the pattern. 
%詳細な議論はDiscussion節で行います。
Detailed discussion of this will be provided in Sec.~\ref{sec_discussion}~Discussion.}
Hereafter, the parameter regions delineated by dotted lines are referred to as pattern states I--VI.\par

By comparing the characteristics of the defined pattern states I--VI and the difference between ${\rm PD}_{\rm long}$ and ${\rm PD}_{\rm short}$, we can attain the physical meanings of the regression results of $\alpha$ and $v$ using part of ${\rm PD}$, that is, ${\rm PD}_{\rm long}$ and ${\rm PD}_{\rm short}$. 
Among the four features that characterize pattern states I--VI, the features that correspond exclusively to ${\rm PD}_{\rm short}$ and ${\rm PD}_{\rm long}$ are the variances of the magnetic moments within the inverse and positive magnetic domains and the Betti number. 
Therefore, ${\rm PD}_{\rm short}$ is a feature that captures the pattern transition around $\alpha=2.0$ caused by the increase in $\alpha$, and ${\rm PD}_{\rm long}$ is a feature that directly captures the sea-island structure caused by the increase in $v$. 
This is consistent with the results of the regression analysis showing that ${\rm PD}_{\rm short}$ was more useful than ${\rm PD}_{\rm long}$ for estimating $\alpha$ and that ${\rm PD}_{\rm long}$ did not perform worse than ${\rm PD}_{\rm all}$ in the $v$ regression. 
Thus, the classification constructed based on TDA was reasonable as a diagram of the state of the magnetic domain structure. 
\par
The obtained classification diagram suggests that there are pattern states I and V with \textcolor{black}{similar} labyrinth patterns but with different properties of magnetic moment intensity fluctuations as $\alpha$ increases. 
%Figs.~\ref{fig7}(a) and (b) show that with increasing $\alpha$, the magnetic domain pattern changes from the labyrinthine structure (I) to a mixture of labyrinthine and sea-island structures (II and IV), and back to the labyrinthine structure (V). 
%It is interesting that with increasing $\alpha$, a magnetic domain pattern with a labyrinthine structure recursively appears and the intradomain fluctuations become larger in the intervening parameter regions. 
%On the basis of the results of the TDA of the pattern formation process in the next section, we will discuss the physical mechanisms behind these labyrinth states. 
Classification by the topological features based on the binarization of the image, i.e., the Betti number, which was used as a feature in previous studies~[2,5], cannot clearly distinguish among pattern states I and V. 
The fact that we were able to classify pattern states I and V as different structures is due to TDA that treats $\phi\left(\bm{r}\right)$ as a continuous variable, rather than a binarized topological feature of the pattern such as the Betti number.\par

\section{Analysis of pattern formation process}
\label{sec_result2}
\begin{figure}[htbp]
  \begin{center}
   \includegraphics[width=\linewidth]{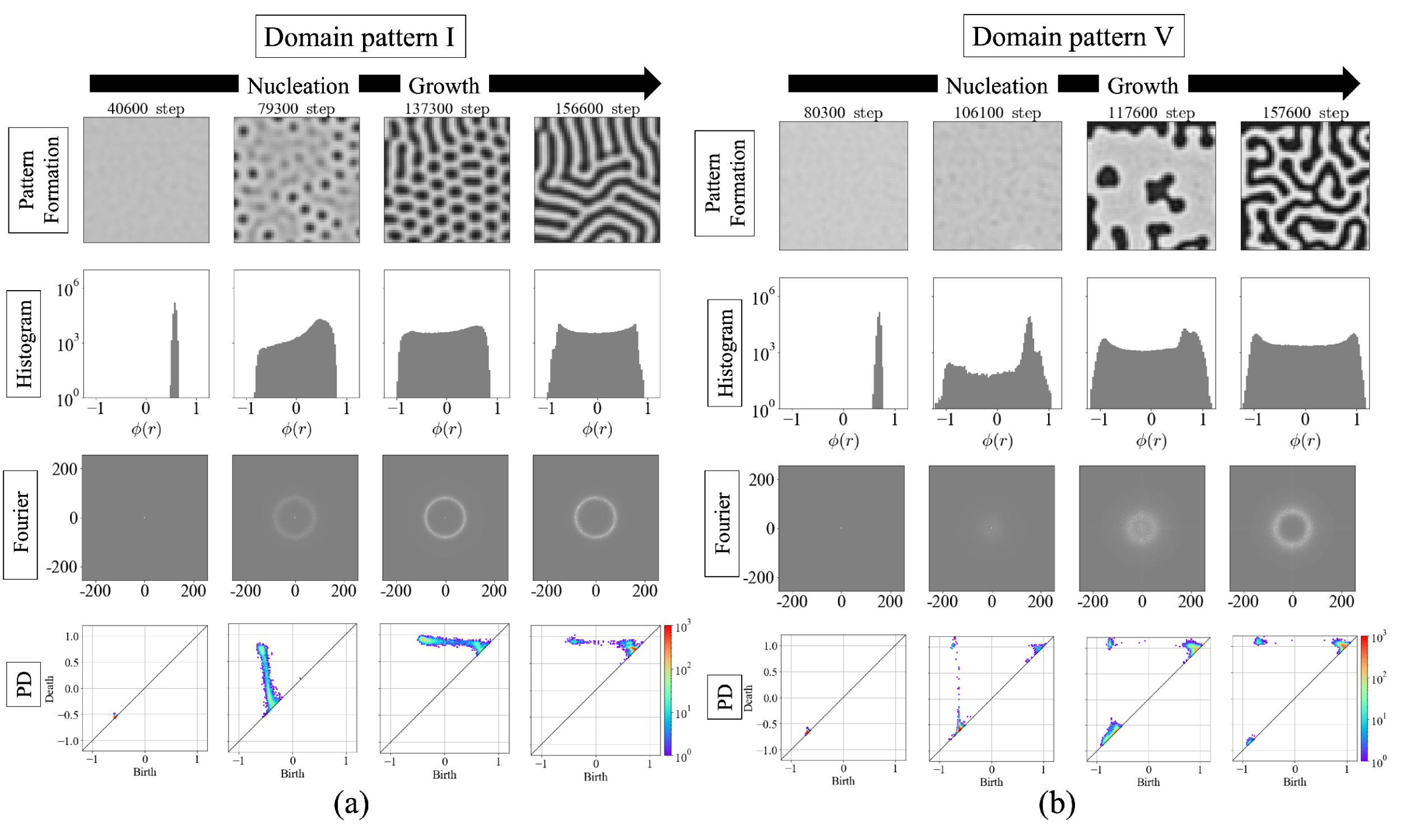}
  \caption{Process of magnetic domain pattern formation and the time evolution of PD.  
  The row showing ``Pattern Formation'' in the figure displays the time evolution of the magnetic domain pattern itself, whereas the row showing ``Histograms'' displays the time evolution of the histograms of the intensity distribution of the magnetic moment $\phi\left(\bm{r}\right)$, and the row showing ``Fourier'' displays the time evolution of the Fourier transformation of the intensity distribution of the magnetic moment $\phi\left(\bm{r}\right)$. The row showing ``PD'' displays the time evolution of the PD. 
  (a) Pattern formation process of labyrinthine structure and the time evolution of the PD in a region with a small $\alpha$ (I).
  (b) Pattern formation process of labyrinthine structure and the time evolution of the PD in a region with a large $\alpha$ (V).}
  \label{fig8}
  \end{center}
\end{figure}
In this section, we elucidate the physical mechanism underlying the classification results in the previous section by TDA of the magnetic domain formation process. 
We focus on the formation processes of pattern states I and V because they have \textcolor{black}{similar} labyrinthine structures but are not adjacent to each other in the parameter space $\alpha$--$v$, which should have not been quantitatively discriminated until the analysis of this study. 
Examples of the magnetic domain pattern formation belonging to pattern states I and V are shown in the first row of Fig.~\ref{fig8}. 
We calculate the PDs of pattern states I and V at each time of the evolution process until the formation of the magnetic domain pattern and analyze the transition of its topological features. 
The transitions of PD according to the formation processes of pattern states I and V are distinctly different. 
In particular, the second and third columns of Figs.~\ref{fig8}(a) and (b) show the time when the inverse magnetic domain is generated and the time when the inverse magnetic domain grows to form a labyrinthine structure, respectively. 
The time when the inverse \textcolor{black}{magnetic} domain is generated is the time when the initial microregions with a value of $\phi < 0$ are generated, and the growth time of the inverse \textcolor{black}{magnetic} domain is the time when the area of the inverse magnetic domain grows (see ``Histogram'' in Fig.~\ref{fig8}). 
The pattern states I and V, which form \textcolor{black}{similar} magnetic domain structures, are found to have very different PD transition processes in their inverse magnetic domain generation and growth processes [see the rows of PD in Figs.~\ref{fig8}(a) and (b)]. 
In other words, the PDs of pattern state I during the generation and growth of inverse magnetic domains have a continuous and elongated distribution structure, whereas the PDs of pattern state V have three separated distributions corresponding to ${\rm PD}_{\rm long}$ and ${\rm PD}_{\rm short}$. \par

\begin{figure}[htbp]
  \begin{center}
   \includegraphics[width=\linewidth]{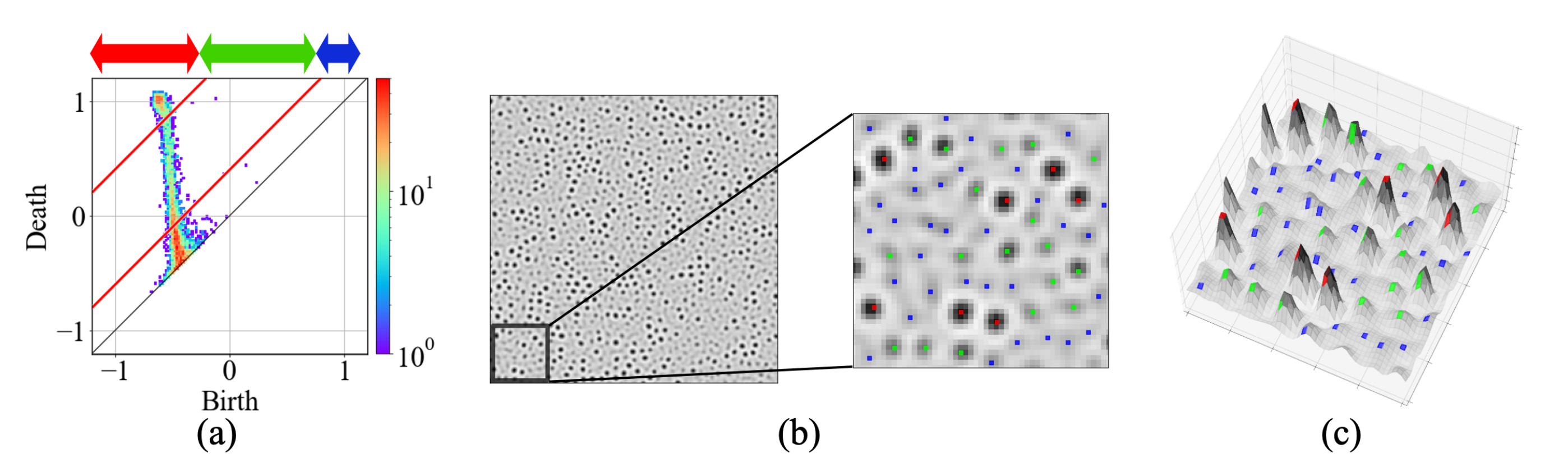}
  \caption{Correspondence between the distribution of generators and the magnetic domain pattern of the PD at the time of domain generation. (a) PD of the magnetic domain pattern ($\alpha =1.5$, $v=0.01$) around the time of domain generation, where the region with long-life generators is indicated by a red arrow, the region with intermediate-life generators is indicated by a green arrow, and the region with short-life generators is indicated by a blue arrow. (b) Magnetic domain pattern and its magnified view. Each color point in the magnified image represents the position of each death cube corresponding to the lifetime region of the corresponding color in the PD. (c) Landscape of magnetic domain pattern, where each color point represents the position of each death cube corresponding to the lifetime region of the corresponding color in the PD. The HomCloud library~\cite{homcloud} was used to extract the locations of death cubes.}
  \label{fig9}
  \end{center}
\end{figure}

The peculiar behavior of the PD at the time of generation of inverse magnetic domains is investigated by inverse analysis. 
The position of the death cube corresponding to the PD of pattern state I  at the time of domain generation is given in Fig.~\ref{fig9}. 
It is found that the continuous distribution structure of PD is due to the fact that the generated domains take not only large positive and small negative values of the magnetic moment but also intermediate values. 
This inverse analysis suggests that, in the parameter region where $\alpha$ is small, the magnetic domains are formed through the continuous increase in the intensity of magnetic moment $\phi\left(\bm{r}\right)$. 
On the other hand, in the region where $\alpha$ is large, the magnetic domains do not take the intermediate intensity of magnetic moments and they emerge discretely. 
\textcolor{black}{
%これは、ドメインが成長する時も同様に考えることができ、pattern states Iでは、ドメインは丸い逆磁区の間の領域が立ち上がってきて逆磁区が連結するようにして成長する。
In pattern state I, domains grow such that the regions between the circular inverse magnetic domains rise up and the inverse magnetic domains are connected (see grayscale image at growth in Fig.~\ref{fig8}(a)). 
%この連続的な立ち上がりに応じて、birth timeが連続的に変化するため、Fig.~\ref{fig8}(a)のPDのような連続的な分布がgrowth時に観測される。
Since the birth time varies continuously with the continuous development of the inverse magnetic domain in this intermediate region, a continuous distribution similar to the PD at growth in Fig.~\ref{fig8}(a) is observed. 
%pattern states Vでは、逆磁区が拡大することで成長するため、growth時にそのような連続的な分布構造を持たない。
In pattern state V, the PD in Fig.~\ref{fig8}(b) during growth does not have such a continuous distribution structure because it grows by expanding the inverse magnetic domain region (see grayscale image during growth in Fig.~\ref{fig8}(b)).
}\par
To investigate how the continuous and discrete magnetic domain generation processes are related to the control parameters $\alpha$ and $v$ of the system, we quantify the maximum ratio of the number of holes that take intermediate states in the process of pattern formation during their generation and growth as the maximum ratios of intermediate state $R^{\rm gen}_{\rm max}$ and $R^{\rm grow}_{\rm max}$ are defined respectively~(detailed definitions of $R^{\rm gen}_{\rm max}$ and $R^{\rm grow}_{\rm max}$ are described in \ref{proc_inter}). 
\textcolor{black}{
%最後に、$\alpha$による変化に注目するために、$\alpha=1.0$で$R^{\rm gen}_{\rm max}$ and $R^{\rm grow}_{\rm max}$が1.0となるように規格化した。
Finally, to focus on changes due to $\alpha$, we normalized $R^{\rm gen}_{\rm max}$ and $R^{\rm grow}_{\rm max}$ to be 1.0 at $\alpha=1.0$.}

\begin{figure}[htbp]
  \begin{center}
   \includegraphics[width=\linewidth]{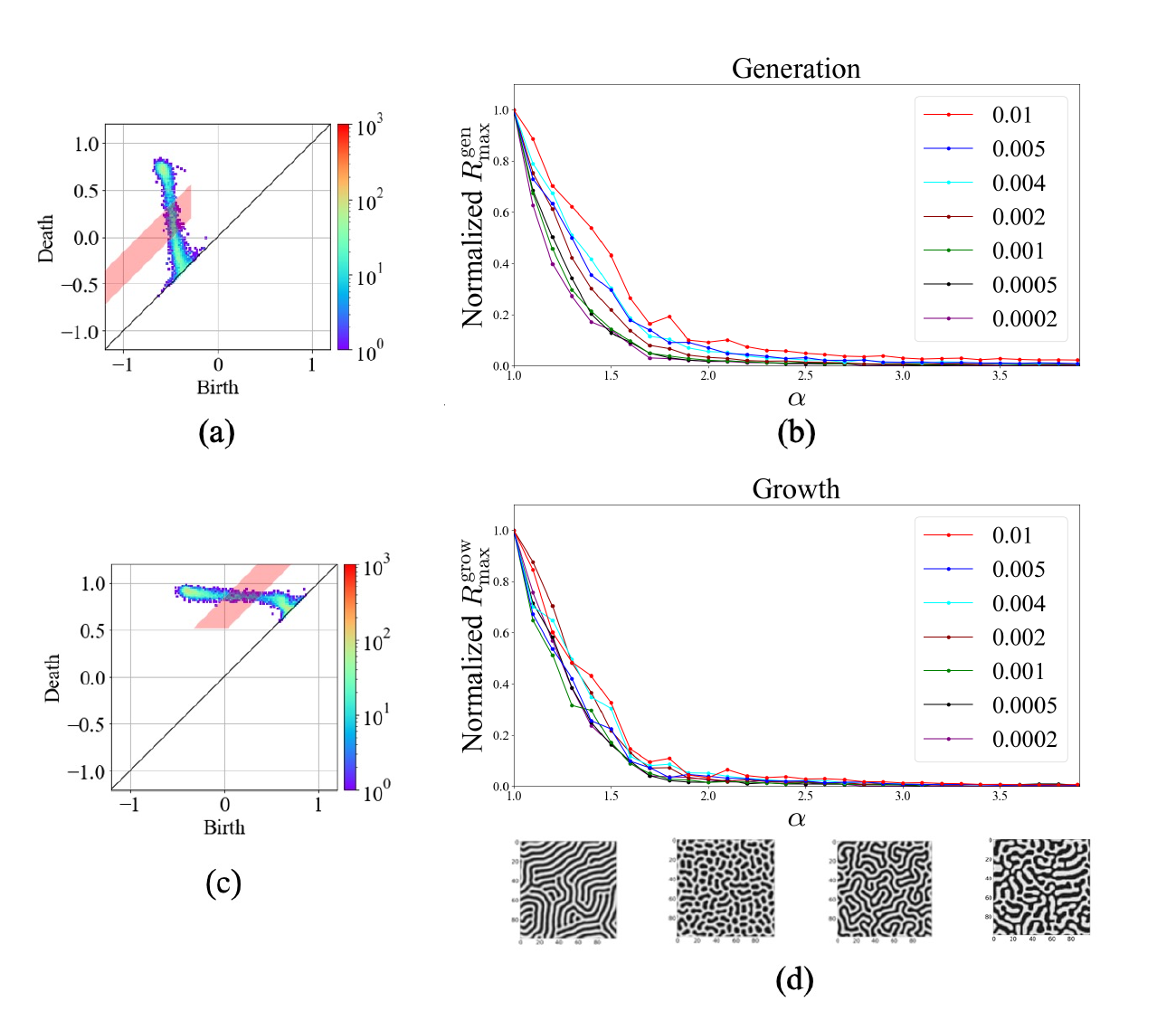}
  \caption{Relationship of the ratio of the number of generators in intermediate state with the total number of generators and $\alpha$ and $v$. (a) PD around the time of domain generation, where the red region is defined as the region of the intermediate state of domain generation. (b) \textcolor{black}{Normalized} ratio of generator number with intermediate state around the time of domain generation. The horizontal axis is $\alpha$, and the vertical axis is the ratio of generator $R^{\rm gen}_{\rm max}$ with intermediate state. (c) PD around the time of domain growth, where the red region is defined as the region of the intermediate state of domain growth. (d) \textcolor{black}{Normalized} ratio of generator number with intermediate state around the time of domain growth. The horizontal axis is $\alpha$, and the vertical axis is the ratio of generator $R^{\rm grow}_{\rm max}$ with intermediate state.}
  \label{fig10}
  \end{center}
\end{figure}

It is confirmed that \textcolor{black}{the normalized $R^{\rm gen}$ at $v<0.005$} and \textcolor{black}{the normalized $R^{\rm grow}$} take relatively large values when $\alpha<2.0$, and decrease rapidly to 0 at around $\alpha=2.0$ [Figs.~\ref{fig10}(b) and (d)]. 
This result indicates that there are two types of magnetic domain pattern formation process for the \textcolor{black}{similar} labyrinthine structures: one involves inverse magnetic domains that were generated and grew discontinuously, and the other involves those that were generated and grew continuously. 
The parameter regions of the first and second types of formation process correspond approximately to the parameter regions of pattern states I and V, respectively. 
In addition, $\alpha=2.0$, where this magnetic domain pattern formation process transitioned, corresponds to the region of pattern states III and IV, which are the sea-island structure and mixed states, respectively.\par

%The result that different magnetic domain formation processes occur rapidly as $\alpha$ increases can be understood from the anisotropic energy $H_{\mathrm{ani}}$, whose relative intensity is controlled by $\alpha$. 
%In other words, since $\alpha$ represents the intensity of magnetic anisotropy, it is easier to form an intermediate state when this value is small, and it is more difficult to form an intermediate state when this value is large. 
%These insights are also obtained owing to the treatment of $\phi\left(\bm{r}\right)$ as a continuous variable in TDA. 
%We can quantitatively determine that there are two types of domain formation process in labyrinthine structures using methods other than feature extraction based on persistent homology, such as analysis of histograms. \par

\section{Discussion}
\label{sec_discussion}
\begin{figure}[htbp]
  \begin{center}
   \includegraphics[width=0.8\linewidth]{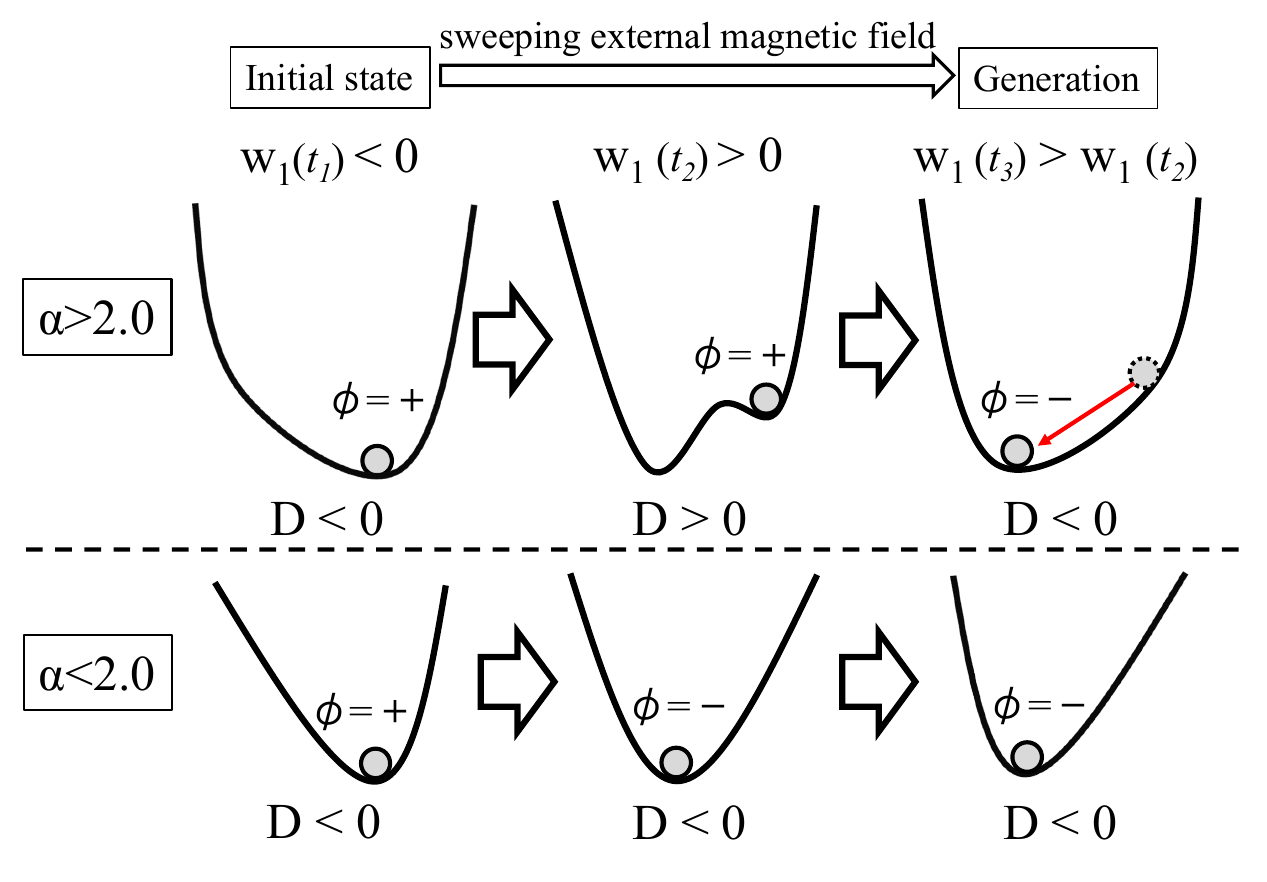}
  \caption{Schematic diagram of the generation mechanism of two different inverse magnetic domains such as pattern states I and V. The time evolution of the reduced energy landscape with the scalar variable $\phi$ is shown. $\phi$ is a representative value of the magnetic moment of the inverse \textcolor{black}{magnetic} domain.}
  \label{fig11}
  \end{center}
\end{figure}

Here, by analyzing the energy function, we discuss why there are two types of magnetic domain pattern formation process for similar labyrinthine structures but have different domain formation processes, which change from continuous to discrete formation depending on the increase in $\alpha$.  
From the results of numerical simulations, it can be seen that when an inverse magnetic domain is generated, the region other than the inverse magnetic domain does not change in response to changes in the external magnetic field, but only the region within the inverse magnetic domain changes (see \ref{appendix2}). 
Therefore, by placing the scalar variable $\phi$ as the magnetic moment characterizing the inverse magnetic domain, the energy function $H(\phi\left(\bm{r}\right))$ [Eq.~\eqref{energy}] is reduced to the following (see \ref{appendix2}): 
\begin{eqnarray}
\textcolor{black}{H}(\phi) = w_0(\beta) + w_1(\beta,\gamma,h(t))\phi + w_2(\alpha,\beta,\gamma)\phi^2  + w_4(\alpha)\phi^4,
\end{eqnarray}
where $w_0$, $w_1$, $w_2$, and $w_4$ represent constants independent of $\phi$. 
If we focus on the case where an isomorphic magnetic domain is generated, we find that $w_1\propto -h(t) +C_1(\alpha)\propto t - C'_1(\alpha)$, $w_2\propto -\alpha + C_2$, $w_4\propto\alpha >0$, and $C_2 > 0$. 
Since only $w_1\phi$ is an asymmetric term of energy function $\textcolor{black}{H}(\phi)$, $w_1<0$ at $t=0$ and changes to $w_1>0$ around the time of domain generation. 
If $\textcolor{black}{H}(\phi)$ takes a sufficiently large $\alpha$ and the system exist around domain generation $w_1\sim 0$, then  we find that $\textcolor{black}{H}(\phi)$ has two minimal solutions from the discriminant formula $D$ for the cubic equation $\frac{d \textcolor{black}{H}(\phi)}{d \phi} = 0$. 
Similarly, the fact that $w_1\propto t$ and $w_1>0$ after the domain generation demonstrates that $w_1$ with a positive value increases with time and that there is a single minimal solution for the cubic equation $\frac{d \textcolor{black}{H}(\phi)}{d \phi} = 0$. 
This indicates that for an energy function with two minimal solutions around the time of domain generation when the value of $\alpha$ is sufficiently large, one of the minimal solutions is eliminated with time. 
Since the TDGL equation is a deterministic time evolution model, the system is expected to remain in a metastable state until the minimal solution of the energy function is retained. 
This type of evolution of an energy landscape at a large $\alpha$ suggests that domain generation passes through supersaturation states ($\alpha>2.0$ in Fig.~\ref{fig11}). 
Similarly, when the value of $\alpha$ is small and there is only one minimal solution even around the time of domain generation, the number of minimal solutions remains to be one despite the increase in time. 
This explains the occurrence of continuous domain generation seen when $\alpha<2.0$ (Fig.~\ref{fig11}). 
This drastic change of the energy landscape with $\alpha$ is the mechanism behind the two similar pattern states I and V with different formation processes. \par
\textcolor{black}{%提案した磁区構造形成ダイナミクスの縮約モデル$\frac{\partial \phi}{\partial t} = \frac{\partial H'(\phi)}{\partial\phi}$を元にした考察で得られた２つのドメイン形成機序の存在は、磁性材料以外のliquid-liquid transition（LLT）と呼ばれる系でも議論されている~\cite{10.1063/5.0021045}。
The existence of two domain formation kinetics as described above has also been discussed in systems called liquid--liquid transition (LLT) other than magnetic materials~\cite{tanaka20, kurita04}. 
%LLTとは、単一成分の物質が2つ以上の液体状態を持つ「液体多形」として知られる系で生じるドメイン生成を伴う遷移過程（LLT）である。
LLT is a transition process with domain formation that occurs in systems known as “liquid polymorphs,” where a liquid with a single-component substance has two or more liquid states. 
%LLTで議論される遷移過程には、nucleation-growth type and spinodal-decomposition typeと呼ばれる２つのタイプのダイナミクスがある。
There are two types of dynamics in the transition process discussed in LLT, called the nucleation-growth type and spinodal-decomposition type. 
%nucleation-growth typeが、本研究における$\alpha>2.0$における不連続な磁区構造形成過程に、spinodal-decomposition typeが$\alpha<2.0$の連続的な磁区構造形成過程に対応する動力学的性質を持つと考えられる~\cite{10.1063/5.0021045}。
The nucleation-growth type should correspond to the discontinuous magnetic domain formation process for $\alpha>2.0$ in this study, whereas the spinodal-decomposition type should correspond to the continuous magnetic domain formation process for $\alpha<2.0$~\cite{tanaka20, kurita04}. 
%このような異なる系における類似したドメイン形成メカニズムの存在は、系に固有の現象論を超えた統一的な視点が得られる可能性を示唆している。
The existence of similar domain formation mechanisms in these different systems suggests the possibility of a unified perspective beyond system-specific phenomenology. 
%また、LLTのデータに本研究の提案枠組みを適用することで、LLTの機序解明が促進されることも期待される。
It is also expected that the application of the proposed framework to LLT will facilitate the elucidation of the kinetics of LLT. 
%本研究で分析対象としたデータはTDGLに基づくシミュレーションデータであるが、LLTの二つの異なるダイナミクスは実際に計測されている。
Although the data analyzed in this study are simulated data based on the TDGL, two different dynamics of LLT have actually been measured~\cite{kurita04}. 
%本研究の結果得られた知見を元に、実現象の動力学モデルの精緻化などをする際のターゲット現象として、LLTに取り組むことを検討している。
Thus, on the basis of the proposed framework and the knowledge about the kinetics of the pattern formation process obtained in this study, we are planning to conduct analytical studies of the pattern formation process in real phenomena or studies to improve the kinetic model for the magnetic domain formation process or LLT. }

%From this reduced energy model, we can also discuss the mechanism of the sea-island state III and mixed states II and IV at around $\alpha = 2.0$. 
%$\alpha \sim 2.0$ should be the boundary parameter region between the continuous and discontinuous magnetic domain generation ($w_2 \sim 0$). 
%Therefore, owing to the spatial inhomogeneity $\lambda$ of anisotropy, some spatial regions are expected to be continuous and others discrete in magnetic domain generation. 
%Compared with the case where magnetic domains are generated continuously, the discrete generation of magnetic domains is relatively delayed because the magnetic domains are not generated until the bimodal structure is resolved. 
%A state in which there is already an inverse magnetic domain can be seen as an additional negative external magnetic field. 
%This might prevent the resolution of the bimodal structure, thus creating a structure that partially lacks the labyrinthine structure. 
%Although we only discussed the domain generation here, similar mechanisms are expected to work in domain growth. 
%Thus, we can confirm that persistent homology is capable of extracting useful features that can reveal the mechanisms of the magnetic domain pattern formation process.\par
Here, we discuss the advantages of analyzing pattern dynamics with persistent homology.
The Fourier spectrum and histograms are considered to be the most commonly used features in pattern data analysis. 
In the case of using features of Fourier spectrum, the peak position of the spectrum directly represents the periodicity of a domain, and the variance of its distribution represents the displacement from a periodic structure such as the curvature of the domain boundary. 
Similarly, even for features used for evaluating the spatial correlation structure of the domain pattern, rotational symmetry, and others, such information about the outline of the domain could be extracted directly from its feature space\cite{mendoza2015nature}. 
For example, the distinction between only the I and V phases of the stable state at time $t=2T_0$ may be captured from the variance of peak structure in the Fourier spectrum resulting from a difference between the curvatures of the domain boundaries (Fig.~\ref{fig8}). 
On the other hand, features such as fluctuations of magnetic moments in the domain, which were found to be important features in this study, and the continuity/discontinuity of magnetic moment values in the domain during the growth process of inverse magnetic domains do not correspond directly to the interpretable feature in the Fourier spectrum. 
Because the fluctuations of magnetic moments in the domain, which could be represented as a high-frequency component of the Fourier spectrum, would not always have a well-defined peak structure, as is the case in this study (Fig.~\ref{fig8}). 
It is generally not easy to extract interpretable information from such a spectrum without assuming a physics model. 
Another commonly used feature space is the feature space with statistics such as histograms, which enables the direct representation of features, such as fluctuations of magnetic moments in the domain and continuous/discontinuous growth of inverse magnetic domains, in this feature space (Fig.~\ref{fig8}). 
On the other hand, the domain's outline information does not correspond directly to its feature space. 
Persistent homology can map both the domain outline information and the statistical information in the domain directly to the feature space. 
That is, persistent homology can map statistical information such as intradomain fluctuations as the spread of generator distributions belonging to ${\rm PD}_{\rm short}$, domain outline information as the number of generators or their birth--death coordinates in ${\rm PD}_{\rm long}$, and information such as continuity or discontinuity in the domain growth process as transitions between these distributions. 
Thus, analysis based on persistent homology can represent a wider range of information than other features in an interpretable form in the feature space. 
This analytical method is expected to work effectively for analyzing unknown phenomena with as few assumptions as possible. 
The results obtained so far in this study suggest that this should work. \par
%In the next section, obtained intuitive understanding of formation process will be discussed more elaborately by analyzing the energy function. \par

\section{Summary} 
\label{summary}
In this study, we have shown that our analysis procedure based on TDA and machine learning techniques could effectively provide important information for quantitatively understanding the magnetic domain formation under a rapidly sweeping external magnetic field on a magnetic thin film with strong uniaxial anisotropy. 
%Using the TDGL equation, we performed numerical experiments on such magnetic domain formations with different anisotropy intensities and sweep rates. 
%Note that, in this study, the other Hamiltonian parameters are fixed. 
%PD represents the hole structures of the magnetic domain pattern as the birth--death of holes according to a change in a threshold value binarizing the grayscale image, i.e., the pattern of magnetic moments $\phi\left(\bm{r}\right)$. 
%Thus, in our analysis, the magnetic moments $\phi\left(\bm{r}\right)$ can be treated as a continuous variable. 
Concretely, by our analysis, we obtained the following findings. 
%In the parameter region evaluated in this study, it is shown that the anisotropy intensities and sweep rates can be estimated using PDs, whose result suggests that PDs retain the information of the magnetic domain formation. 
Using only the short lifetime region of PD, which corresponds to the intensity fluctuations of magnetic moments within each domains, the anisotropy intensities and sweep rates can be estimated precisely. 
This result suggests that not only the global geometric structure of the magnetic domains, as used in previous research, but also the contour features of the magnetic domains, such as fluctuations within the domain, contain information about the physical mechanisms of magnetic domain formation. 
Our procedure can discriminate two labyrinth patterns quantitatively that resemble each other but in different pattern state, that is, as anisotropy intensity continuously increases, the magnetic domain pattern changes from a labyrinthine pattern state to a sea-island pattern state or a mixture of the two, and then back to a labyrinthine pattern state. 
By constructing a reduced Hamiltonian of the system based on analysis results of PDs, we found that such two different labyrinthine pattern states could be understood as the difference between transitions of the energy landscape, that is, continuous and discrete transitions. 
These findings also suggest to experimental researchers that it is important to measure magnetic moment rigorously enough to extract fluctuations of magnetic moments within a domain or the continuity of time evolution of magnetic moments.\par
The analysis procedure using PD presented in this paper is expected to give one format applicable to extracting scientific knowledge from a wide range of pattern formation processes with long-range interactions. 
The data analyzed in this study are just the simulation data of time evolution at $0$ K of the magnetic moment distribution on a magnetic thin film under a decaying external magnetic field that takes values only in the perpendicular direction of the thin film. 
By introducing vector magnetic moments into the simulation to reproduce magnetic wall effects\cite{barbara1994magnetization, saratz2016critical} or by introducing temperature effects into the simulation to reproduce thermodynamic properties\cite{pighin2012finite}, it would be possible to obtain analysis data that reproduce a wider range of real-world magnetic materials. 
Analysis data sets can also be constructed from actual measurement data. 
It has been reported that pattern states, such as the mixture of labyrinth and sea-island pattern states successfully classified in this study, emerge in the simulations of more complex models such as those described above\cite{barbara1994magnetization, saratz2016critical, pighin2012finite} or in actual measurement data\cite{saratz2010experimental, kronseder2015real, novakovic2020stripes, moon2021universal}. 
Therefore, it is expected that the proposed procedure will be effective for a wide range of formation processes for the magnetic domain. 
Verification of the validity of the proposed procedure for a wide range of such practical models or actual measurement data is a subject of our future work.
%\par
%PD-based features should be useful not only for solving the TDGL equation, but also for the analysis of systems in the continuum approximation with similar energy functions and the corresponding measurement data. 
%Furthermore, because PD can be used to estimate physical properties and analyze time-series data, it is possible to develop a prediction model of the time evolution of physical properties using the time evolution model of PD. 

\appendix

\section{Detailed numerical calculation methods of TDGL equation}
\label{appendix_TDGL}
By performing variational differentiation [Eq.~\eqref{eq_variation}], we obtain the time derivative of $\phi(\bm{r})$ as
\begin{eqnarray}
\label{eq_functional}
\frac{\partial\phi(\bm{r})}{\partial t} = \alpha \lambda(\bm{r})\left[\phi(\bm{r}) - \phi(\bm{r})^3\right] + \beta \nabla^2\phi(\bm{r}) - \gamma \int d\bm{r}'\phi(\bm{r}') G(\bm{r}, \bm{r}') + h(t).\nonumber  \\
\end{eqnarray}
Fourier transformation of Eq.~\eqref{eq_functional} provides another perspective of the equation as follows:
\begin{eqnarray}
\label{eq_k}
\frac{\partial\tilde{\phi}_{\bm{k}}}{\partial t} = \alpha \left\{\tilde{\phi}(\bm{k}) \star \tilde{\lambda}(\bm{k}) - [(\tilde{\phi}(\bm{k}) \star \tilde{\phi}(\bm{k})) \star \tilde{\phi}(\bm{k})] \star \tilde{\lambda}(\bm{k})\right\} - [\beta k^2 - \gamma \tilde{G}({\bm{k}})]\tilde{\phi}(\bm{k}) + h_{\bm{k}} \delta_{{\bm{k}},0},\:\:\:\:\:
\end{eqnarray}
where $\tilde{\phi}(\bm{k})$, $\tilde{\lambda}(\bm{k})$, and $\tilde{G}({\bm{k}})$ denote the Fourier transforms of $\phi(\bm{r})$, $\lambda(\bm{r})$, and $G(\bm{r}, 0)$ respectively, $\star$ denotes the convolution sum, and $k=|{\bm{k}}|$. 
In addition, the following conversion equation is used:
\begin{eqnarray}
\int d\bm{r} \int d\bm{r}'\phi(\bm{r}') G(\bm{r}, \bm{r}') \exp(-i\bm{k}\cdot\bm{r})=\tilde{G}({\bm{k}})\tilde{\phi}(\bm{k}).
\end{eqnarray}
Specifically, $\tilde{G}({\bm{k}})$ is given by the following equation:
\begin{eqnarray}
\tilde{G}({\bm{k}}) = a_0 - a_1 k,
\end{eqnarray}
where $a_0=2\pi \int_{d}^{\infty} dr \frac{1}{r^2} = \frac{2\pi}{d}$ and $a_1 = 2\pi$. 
The time evolution [Eq.~\eqref{eq_k}] is achieved as a difference equation with a time increment of $0.1$. 
In the numerical calculations, the space $\bm{r} := (x,y)$ is discretized into a mesh. 
Therefore, note that the distribution function of the average magnetic moment $\phi(\bm{r})$ obtained from the numerical calculations is given as $\phi(i,j), \:i,j\in \mathbf{N}$.\par

\section{Parameter dependency of TDGL equation}
\label{appendix_TDGL_depend}
The TDGL equation changes the formation pattern depending on the model parameters $\alpha$, $\textcolor{black}{\nu}$, $\beta$, $\gamma$, and $v$. 
To investigate the behavior of the TDGL equation with respect to the parameters, the linear amplification factor $\eta_{\bm{k}}$ at zero magnetization $h_{\bm{k}}=0$ of Eq.~\eqref{eq_k} was previously studied~\cite{kudo2007field}. 
From Eq.~\eqref{eq_k}, the system's linear amplification factor is given by
\begin{eqnarray}
\frac{\partial\tilde{\phi}_{\bm{k}}}{\partial t} = \eta_{\bm{k}} \tilde{\phi}_{\bm{k}},
\end{eqnarray}
\begin{eqnarray}
\label{eq_linear_increase}
\eta_{\bm{k}} &=& -(\beta k^2 - \gamma a_1 k + \gamma a_0) + \alpha\\
&=& -\beta \left(k - \frac{a_1\gamma}{2\beta}\right)^2+\frac{a_1^2\gamma^2}{4\beta} - \gamma a_0 + \alpha.
\end{eqnarray}
From these equations, the wavenumber component at $k = \frac{a_1\gamma}{2\beta}$ is the largest component, and the larger $\frac{a_1^2\gamma^2}{4\beta} - \gamma a_0 + \alpha$ becomes, the broader the wavenumber component becomes. 
This indicates that $\frac{a_1\gamma}{2\beta}$ determines the rough scale of the domain structure, and $\alpha$ controls the variability of the domain size. 
This signifies that even if $\beta$ and $\gamma$ are fixed, we can still generate some variety of patterns by controlling $\alpha$ in the range that can be explained by linear amplification without an external magnetic field. 
From this consideration, Kudo and Nakamura~\cite{kudo2007field} fixed $\frac{a_1\gamma}{2\beta}$, which specifies the scale of the domain structure, to $1$ and focused on the $\alpha$ dependence of the domain structure. 
Kudo and Nakamura also focused on the $v$-dependence controlling the nonequilibrium effects of the domain structure formation dynamics.

\section{Differences in persistent homology groups due to variations in the construction method of a cubical complex}
\label{appendix}
The term ``holes'' has been used ambiguously in the main text. 
Depending on how holes are quantified from the figure data, different persistent homology groups are obtained  from the same data. 
This section explains how the construction method of a cubical complex produces different persistent homology groups. 
For this purpose, the calculation procedure for PD is explained in detail. 
The characteristics of each persistent homology group obtained by the two libraries with different construction methods employed in this study, GUDHI~\cite{dlotko2018computational} and  HomCloud~\textcolor{black}{\cite{homcloud}} library, are described in the last paragraph of this appendix.\par
To quantitatively establish holes, figure data are first reexpressed as a set of triangles (a simplicical complex) or rectangles (a cubical complex) to ensure that they can be dealt with algebraically. 
When grayscale pixel image data $G:=\{G(i,j)\in [0,1]\mid\: 1\leq i \leq \textcolor{black}{p}_x, 1\leq j \leq \textcolor{black}{p}_y\}$ are provided as input data, they are converted to a cubical complex using the following data structure: we consider grayscale pixel data as a set of binary images binarized at a certain threshold $t$, which is expressed as
\begin{eqnarray}
G &=& \{G^{\rm binary}_{t} \mid t\in[t_{\rm min},t_{\rm max}]\},\\
G^{\rm binary}_{t} &=& \{G(i,j)\mid G(i,j)\geq t\},
\end{eqnarray}
and represent each binary image $G^{\rm binary}_{t}$ as a cubical complex $K^t$. 
A pixel in a binary image $G^{\rm binary}_{t}$ is considered a rectangle, and the structure of this image can be covered by rectangles [Fig.~\ref{fig2}(b), upper panel]. 
Because a rectangle is a closed convex set, this coverage can be considered a representation of a figure by a set of closed convex sets, which can be transformed into a cubical complex with the same topological structure. 
For example, by mapping a convex set (pixel) to a cube, a cubical complex can be constructed as follows [Fig.~\ref{fig2}(c), upper panel]: If a pixel $G(i,j)$ belongs to a domain, then a zero-dimensional cube denoted as $[i,i]\times[j,j]$, called a zero-cube or point, is placed at the center of a rectangle. 
If a neighboring pixel $G(i,j)$ of $G(i\pm1,j)$ or $G(i,j\pm1)$ belongs to a domain, then we assume that $G(i,j)$ and other neighboring pixels are connected and set a one-dimensional cube, called a one-cube or line, which is represented as $[i,i\pm1]\times[j,j]$ or $[i,i]\times[j,j\pm1]$. 
Similarly, if four adjacent pixels $G(i,j)$, $G(i+\delta_i, j)$, $G(i, j+\delta_j)$, and $G(i+\delta_i, j+\delta_j)$, where $\delta_i=\pm1$ and $\delta_j=\pm1$ belong to a same domain, we set a two-dimensional cube, called a two-cube or square, which is denoted as $[i,i+\delta_i]\times[j,j+\delta_j]$. 
For pixel data with $d$ dimensions ($d\geq 3$), we can also set a higher-order cube $[i_1,i_1+\delta_{i_1}]\times[i_2,i_2+\delta_{i_2}]\times\cdots \times [i_d,i_d+\delta_{i_d}]$. 
By appropriately combining the above-mentioned squares, $G^{\rm binary}_{t}$ is represented as a set of cubes or cubical complexes $K_t$. 
The cube elements of $K_t$ increase as the threshold $t$ decreases. 
Here, the threshold value $t$ is replaced by $t'=1-t$ to allow the increase in threshold value to correspond to the increasing sequence of cube elements. 
The sequence of cube elements of $K_{t'}$ is created according to an increase of threshold $t'$ and forms a monotonically increasing sequence. 
This increasing sequence is called a filtration $\mathbf{K}$. 
\begin{equation}
\mathbf{K} := K_{t_1} \subset K_{t_2} \cdots \subset K_{t_{n_{\Theta}}}
\end{equation}
where $0\leq t'_k < t'_{k+1}\leq 1$ and $n_{\Theta}$ is the total number of thresholds $t'$ at which a cube belonging to $\mathbf{K}$ is generated depending on the increase in $t'$. 
To capture the change in graphic structure due to threshold $t'$, we consider the following vector in an $n_{\Theta}$-dimensional space:
\begin{equation}
\label{basis}
\hat{\sigma}_m = (0,0,\ldots,0,\sigma_m,0,\ldots,0),
\end{equation}
where the non-zero element $\sigma_m$ is located at the $n_{b_m}$-th element of the vector, and the $n_{b_m}$-th element corresponds to threshold $b_m$ at which a certain cube $\sigma_m$ generated. 
\textcolor{black}{Note that in this section, $\sigma$ does not represent the kernel width for the vectorization of PD.}
Consider an operator $x$ that acts on this vector space and shifts one element to the right (i.e., moves to the threshold where the next young cube is generated) as follows: 
\begin{equation}
x (c_1,c_2\cdots) = (0,c_1,c_2,\ldots).
\end{equation}
Thus, all cube elements in each cubical complex corresponding to $\mathbf{K}$, which vary with threshold $t'$, can be represented using the basis vector in Eq.~(\ref{basis}) and manipulate using the operator $x$. 
This representation allows for the algebraic manipulation of geometric elements of the figure, such as cubical complexes. \par
Geometrical structures, such as cycles and boundaries, can also be defined and manipulated algebraically by representing them in the same manner. 
Here, we simplify a figure as a binary image and explain the concepts of a cycle, boundary, and hole by assuming that the vector of the cubical complex is one dimensional, i.e., $\hat{\sigma}_m = \sigma_m$. 
This corresponds to the definition of a cycles and boundaries for a binary image in the homology group. 
\begin{figure}[htbp]
  \begin{center}
   \includegraphics[width=\linewidth]{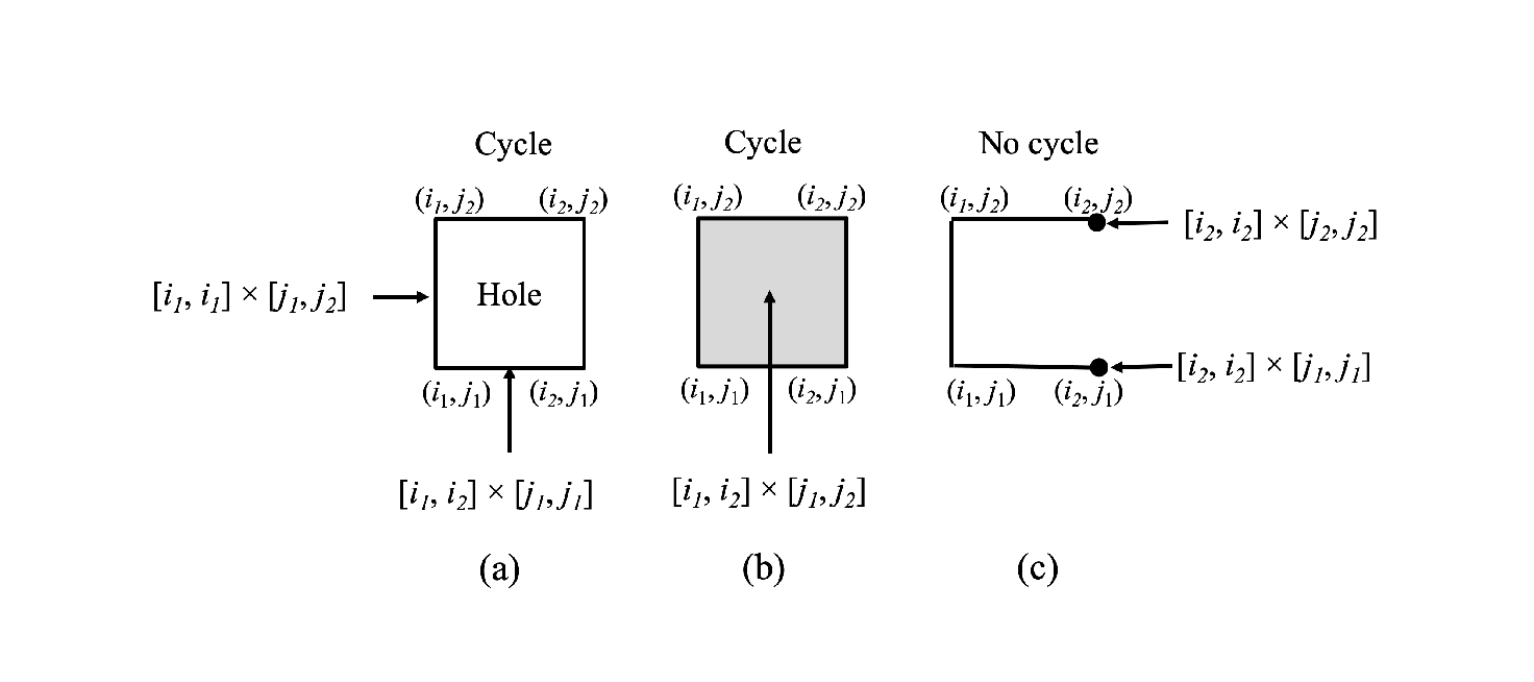}
  \caption{Examples of cubical complexes with (a) a cycle and hole, and (b) a cycle filled by a two-cube, and (c) a one-cubical complex with a two-boundary structure. The boundaries comprise the endpoints of the cubical complex.}
  \label{fig_append0}
  \end{center}
\end{figure} 
A $d$-dimensional cycle, which is the candidate of a hole, is a $d$-dimensional cubical complex that has a structure without boundaries [Figs.~\ref{fig_append0}(a) and (b)]. 
The boundary of a cubical complex comprises its endpoints. 
For example, if a cubical complex is a line (one-cube) $[i_1,j_1]\times [i_2,i_2]$ where $i_1 \neq j_1$, then both endpoints (zero-cubes) of the line, $[i_1,i_1]\times [i_2,i_2]$ and $\:[j_1,j_1]\times [i_2,i_2]$, form its boundary. 
If a cubical complex is a rectangle (two-cube) $[i_1,j_1]\times [i_2,j_2]$ where $i_1 \neq j_1$ and $i_2 \neq j_2$, then its boundary comprises four lines (one-cubes) around a two-cube: $[i_1,j_1]\times [i_2,i_2],\:[i_1,j_1]\times [j_2,j_2],\:[i_1,i_1]\times [i_2,j_2]$, and $[j_1,j_1]\times [i_2,j_2]$. 
Thus, a boundaryless connection signifies that the endpoints of each cube are connected without excess or deficiency, such as $[i_1,j_1]\times [i_2,i_2],\:[i_1,j_1]\times [j_2,j_2],\:[i_1,i_1]\times [i_2,j_2]$, and $[j_1,j_1]\times [i_2,j_2]$. 
Specifically, to be a cycle, there should be exactly two multiples of the endpoint structures of the shape when adding the endpoint structures of each side. 
This is expressed as follows:
\begin{eqnarray}
&\:&([i_1,i_1]\times [i_2,i_2] + [j_1,j_1]\times [i_2,i_2])+([j_1,j_1]\times [i_2,i_2]+[j_1,j_1]\times [j_2,j_2])\nonumber\\
&+&([j_1,j_1]\times [j_2,j_2]+[i_1,i_1]\times [j_2,j_2])+([i_1,i_1]\times [j_2,j_2]+[i_1,i_1]\times [i_2,i_2])\nonumber\\
&=& 2([i_1,i_1]\times [i_2,i_2]) + 2([j_1,j_1]\times [i_2,i_2])+2([j_1,j_1]\times [j_2,j_2])+2([i_1,i_1]\times [j_2,j_2]).\nonumber \\
\end{eqnarray}
Thus, the condition for a given cubical complex to constitute a cycle is that the coefficients resulting from the addition of the boundary structures are multiples of 2. 
To provide a more explicit indicator of the presence or absence of such a cycle, we replace the coefficient from adding the cube complexes corresponding to the endpoints with the remainder of dividing that coefficient by 2, i.e., the coefficient is replaced from integer $\mathbf{Z}$ to factor ring $\mathbf{Z_2}$. 
Consequently, the addition of zero-cubes corresponding to endpoints of one-cubes is expressed as follows:
\begin{eqnarray}
&\:&2([i_1,i_1]\times [i_2,i_2]) + 2([j_1,j_1]\times [i_2,i_2])+2([j_1,j_1]\times [j_2,j_2])+2([i_1,i_1]\times [j_2,j_2])\nonumber\\
&=&0([i_1,i_1]\times [i_2,i_2]) + 0([j_1,j_1]\times [i_2,i_2])+0([j_1,j_1]\times [j_2,j_2])+0([i_1,i_1]\times [j_2,j_2])\nonumber\\
&=&0.
\end{eqnarray}
Thus, finding a cycle corresponds to finding structures in which the boundaries of a cubical complex are added and their coefficients are zero. 
A d-dimensional hole is defined as a $d$-dimensional cycle that is not filled by a (d+1)-dimensional cubical complex, i.e., a $d$-dimensional cycle that is not the boundary of a (d+1)-dimensional cubical complex [Fig.~\ref{fig_append0}(a)]. 
Therefore, to extract the subset of a cubical complex corresponding to a hole, we take the difference (quotient set) between the subset of a $d$-dimensional cubical complex representing d-dimensional cycle structures and the set of a $d$-dimensional cubical complex that forms the boundary of a ($d+1$)-dimensional cubical complex.\par
For a grayscale image, if the threshold $t'$ at which a boundary is generated or disappears is known, the threshold at which hole $k$ is generated $b_k$ and disappears $d_k$ can be calculated. 
The PD can be calculated if the pairs of $b_k$ and $d_k$ for all holes are obtained. 
For this purpose, a boundary is treated in the vector space in the same way as  $\hat{\sigma}_m$. 
The boundary $\hat{\sigma}^l_m$, which is the boundary of an $l$-dimensional cube generated at $m$, is obtained as
\begin{eqnarray}
\partial_l\left(\hat{\sigma}^l_m\right)&=&\partial_l\left(0,0,\ldots ,[i_1,j_1]\times[i_2,j_2]\times\cdots \times [i_l,j_l],\ldots,0,0 \right)\\
&:=& \sum_{k=1}^{l}\left[ x^{b(\sigma^l_m) - b(\sigma^{l-1}_{i_k})} \left(0,0,\ldots,\sigma^{l-1}_{i_k},\ldots,0,0\right)\right. \nonumber\\
&+& x^{b(\sigma^l_m) - b(\sigma^{l-1}_{j_k})}\left.\left(0,0,\ldots,\sigma^{l-1}_{j_k},\ldots,0,0\right)\right],\\
\sigma^{l-1}_{i_k}&:=&[i_1,j_1]\times\cdots\times[i_k,i_k]\times\cdots \times [i_l,j_l],\\
\sigma^{l-1}_{j_k}&:=&[i_1,j_1]\times\cdots\times[j_k,j_k]\times\cdots \times [i_l,j_l],
\end{eqnarray}
where $\partial_l\left(\hat{\sigma}^l\right)$ is a vector whose elements correspond to the threshold at which the boundary is generated, $i_s \neq j_s$ for all $s$, and $b(\cdot)$ represents the index of vector $\hat{\sigma}_m$ [Eq.~\eqref{basis}], and corresponds to the threshold at which the $l$-dimensional cube, $\sigma_m$, generates. 
$\partial_l\left(\hat{\sigma}^l\right)$ is called a boundary operator. 
In the case of a binary image, the boundary set of cubical complex that add to zero correspond to a cycle. 
When adding the vector elements in $\partial_l\left(\hat{\sigma}^l\right)$, each vector element is multiplied by the exponentiation of $x$ to align the vector elements. 
Thus, in the case of a grayscale image, a cycle that is not a boundary is calculated as the case of the binary image. 
Accordingly, the persistence diagram $PD(\mathbf{K})$ specifying the persistent homology group is calculated algebraically in the vector space. 
Thus, once pixel image data are covered by convex sets, the persistent homology group can be calculated.\par
Here, we explain the changes in the calculated persistent homology groups due to the difference in the methods used to cover pixel image data by convex sets. 
One approach to covering pixel image data $G_{t'}(i,j)$, which are binarized at a certain threshold $t'$, is to cover one pixel as a single cube, as described above. 
This coverage treats adjacent diagonal pixels as if they are disconnected. 
Another type of relationship can be considered as a connected cubical complex by setting zero-cubes (points) at the center of a pixel, its four corners, and its four border lines [Fig.~\ref{fig2}(b), upper panel]. 
In this study, the latter type of coverage was chosen because it can extract more diverse information about an object. 
The calculation of persistent homology groups based on such coverage was implemented in the GUDHI library~\cite{dlotko2018computational}. 
In this study, the HomCloud library~\cite{homcloud} was also used to extract the pixels corresponding to the center of a death cube to understand the PDs. 
The HomCloud library uses a coverage method that sets a single zero-cube at the center of a pixel. 
Therefore, it should be noted that in the analysis of the location of a death cube, as described in Sec.~\ref{result_stable}, not all rectangle locations obtained in the GUDHI library were extracted; instead, only their approximate properties were derived.\par

\begin{figure}[htbp]
  \begin{center}
   \includegraphics[width=\linewidth]{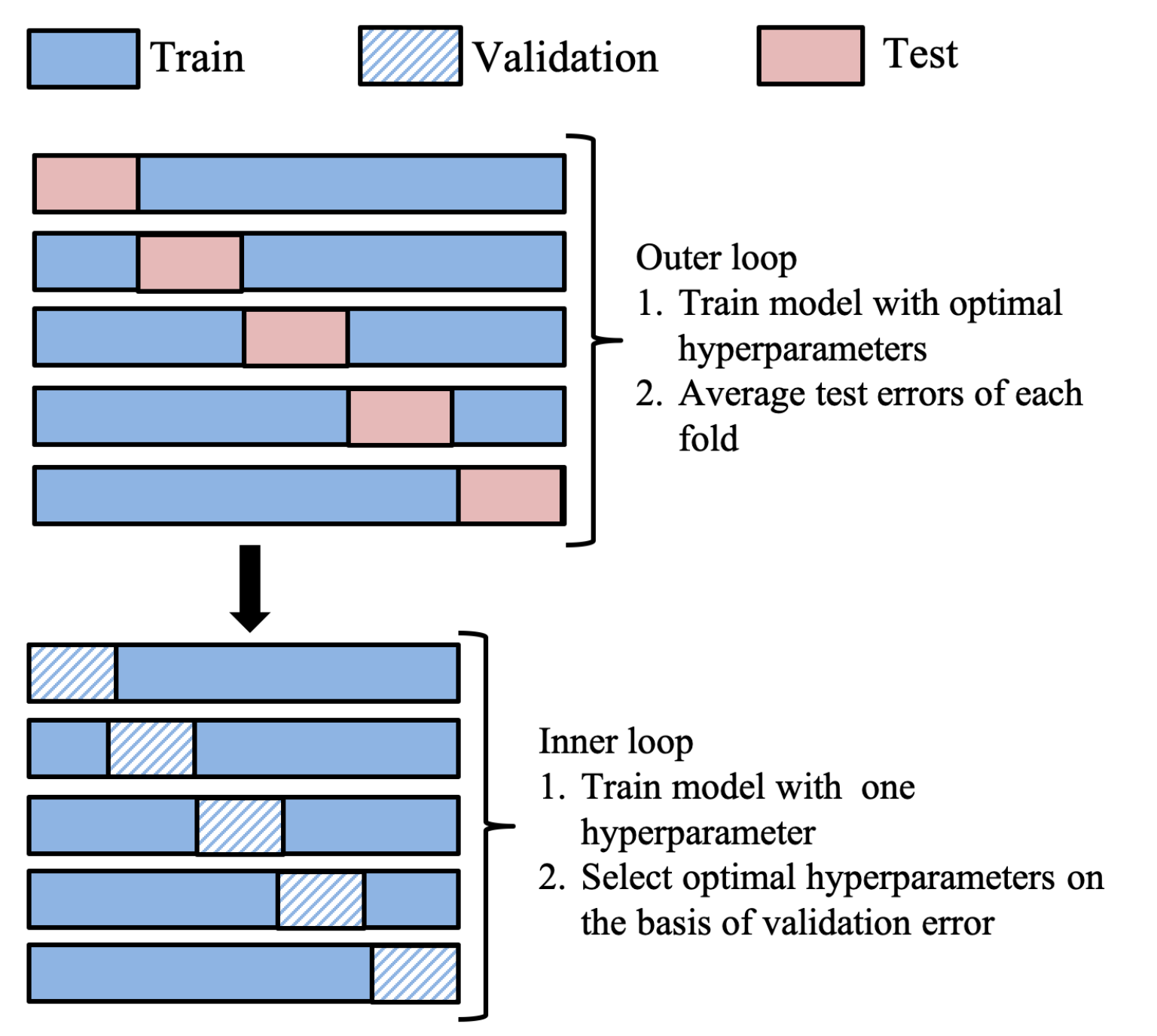}
  \caption{Procedure for the nested cross-validation method~\cite{cawley2010over}.}
  \label{fig_nested}
  \end{center}
\end{figure}

\section{Detailed procedure for the regression analysis}
\label{sec_nested}
For the multiple regression analysis, we performed Ridge regression using the error function.
\begin{eqnarray}
E &=& \frac{1}{2}\sum_{i=1}^{N_{\rm train}}\left[F({\bf v}_i(\sigma,\textcolor{black}{M});\boldsymbol{w}) - y_{i}\right]^2 + \lambda \boldsymbol{w}^T\boldsymbol{w},\\
&\:&y_i \in \left\{\alpha_i,\log\:v_i\right\},
\end{eqnarray}
where $F(\:\cdot\:;\boldsymbol{w})$ is the linear regression function $\mathbf{R}^{d^2}\rightarrow \mathbf{R}$, 
$\boldsymbol{w} := (w_1,w_2\cdots w_{d^2})$ is the regression coefficient, $\lambda$ is the regularization parameter, and $N_{\rm train}$ is the number of training data. 
$\lambda$, the kernel width $\sigma$, and the number of grids $\textcolor{black}{M}$ in vectorizing the PD are hyperparameters of this regression model. 
In this study, we applied the nested cross-validation framework~\cite{cawley2010over} to estimate hyperparameters and generalization errors from a small number of data. \par
Nested cross-validation estimates the generalization error of the underlying model and its hyperparameters based on two cross-validations in a nested relationship (i.e., an inner loop and an outer loop). 
The detailed procedures for both loops are described in Fig.~\ref{fig_nested}. 
In the inner loop, the model is trained with one hyperparameter, and the optimal hyperparameters are selected based on the validation error. 
In the outer loop, the model is trained with optimal hyperparameters obtained from the inner loop, and the prediction error of the model is the average of the test errors of each fold.\par
The regression model is trained using the hyperparameters selected in the inner loop of nested cross-validation, and the generalization error of the model is estimated using the outer loop. 
By plotting together the test predictions $y_i$ for all outer loop folds, we obtain prediction values of all data in $\{{\bf v}_i(\sigma,\textcolor{black}{M}), y_i\}_{i=1}^{96}$ (Fig.~\ref{fig5}). 

\section{Additional analyses for magnetic-domain patterns at $t=2T_0$}
\label{appendix3}

\begin{figure}[htbp]
  \begin{center}
   \includegraphics[width=\linewidth]{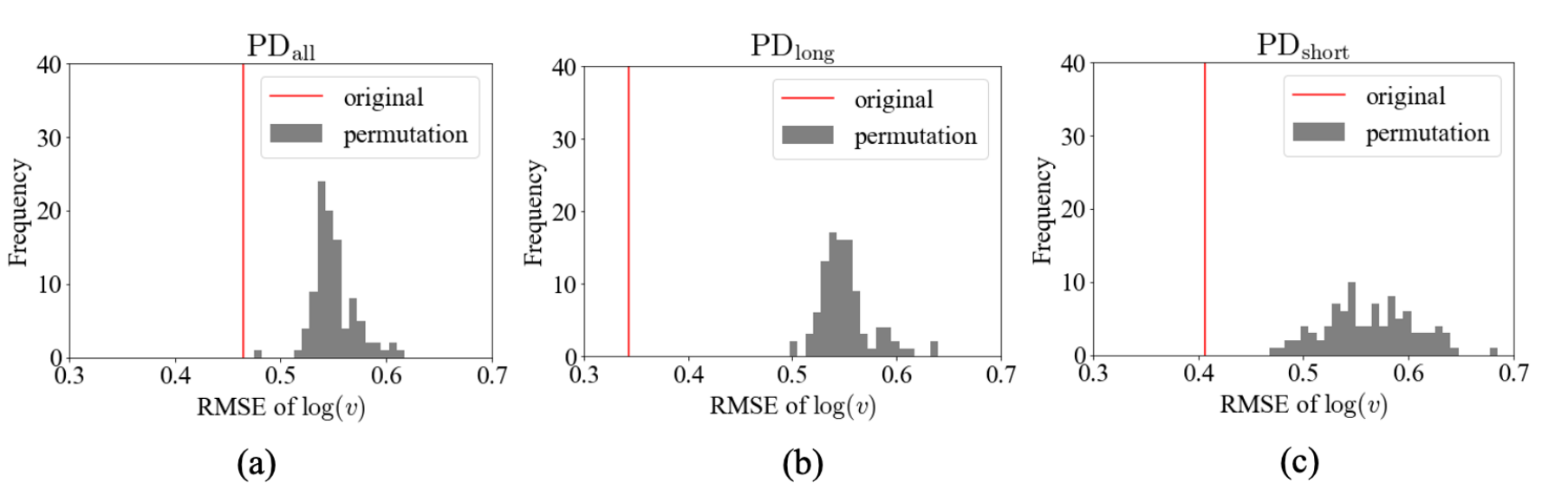}
  \caption{Results of the permutation test for (a) ${\rm PD}_{\rm all}$, (b) ${\rm PD}_{\rm long}$, and (c)  ${\rm PD}_{\rm short}$.}
  \label{fig_append3}
  \end{center}
\end{figure}

\begin{figure}[htbt]
  \begin{center}
   \includegraphics[width=0.75\linewidth]{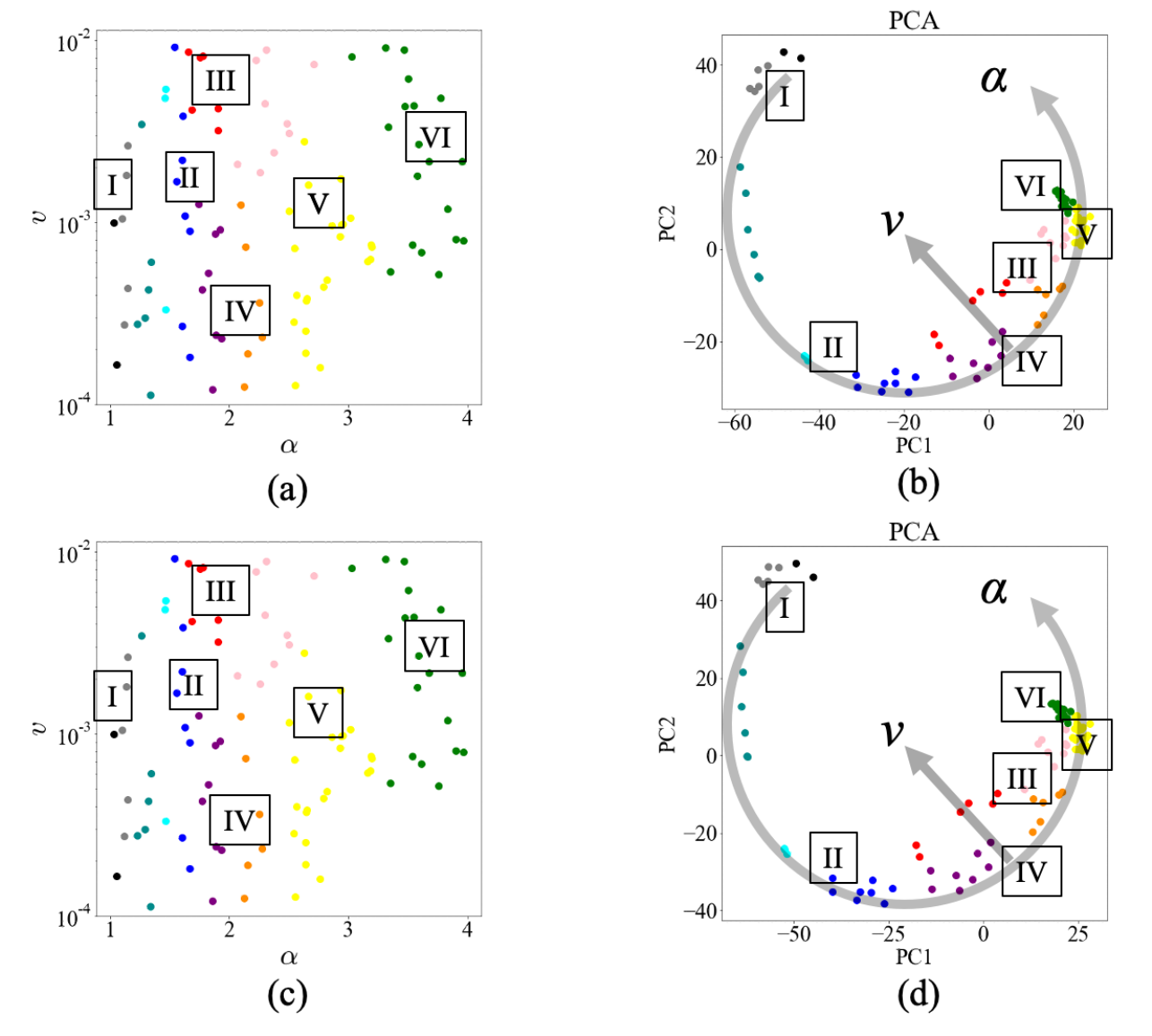}
  \caption{Clustering results of the K-means method and results of the PCA. 
  (a) K-means clustering results of the PDs vectorized with parameters $\sigma$ and $\textcolor{black}{M}$ estimated by regression analysis of $\alpha$. 
  (b) Results of the PCA of the PDs vectorized with parameters $\sigma$ and $\textcolor{black}{M}$ estimated by regression analysis of $\alpha$. 
  (c) K-means clustering results of the PDs vectorized with parameters $\sigma$ and $\textcolor{black}{M}$ estimated by a regression analysis of $\log\:\:v$. 
  (d) Results of a PCA of the PDs vectorized with parameters $\sigma$ and $\textcolor{black}{M}$ estimated by a regression analysis of $\log\:\:v$.}
  \label{fig_append4}
  \end{center}
\end{figure}

Three additional analyses were conducted to support the conclusions drawn from the analysis of a magnetic domain pattern at $t=2T_0$ discussed in the main text. 
The first was a statistical validation that suggests that the PDs are a good feature for estimating $\log\:v$. 
Under the ridge regression of $\log\:v$ with nested cross-validation that is described in the previous section (\textcolor{black}{see} ~\ref{sec_nested}), we statistically tested whether the PDs contain the information required to estimate $\log\:v$. 
In the verification, the RMSE distribution by permutation of the objective variable of the regression, $\log\:\:v$, was calculated and used to the statistical test whether the regression accuracy of the original data was significantly high. 
The results of this permutation test confirmed that for all ${\rm PD}$, the estimation accuracy of $\log\:v$ of the original data was significantly higher than that of the permutated data (Fig.~\ref{fig_append3}). 
These results suggest that the vectorized data of the three parts of the ${\rm PD}$ are good features for estimating $\log\:v$. \par
The second additional analysis supported the K-means clustering results based on ${\rm PD}_{\rm all}$. 
The optimal hyperparameters $\sigma_{\rm opt}$ and $\textcolor{black}{M}_{\rm opt}$ take different values between the regression analysis results of $\alpha$ and $v$. 
We performed clustering analysis for each hyperparameter. 
Since the results were almost identical, in main text, we only show results of clustering using hyperparameters optimized by regression analysis of $\alpha$. 
In this section, we show these results in Fig.~\ref{fig_append4}.
The exact same clustering results and the similar PCA results in the case of $\alpha$ were obtained.\par
\begin{figure}[htbp]
  \begin{center}
   \includegraphics[width=\linewidth]{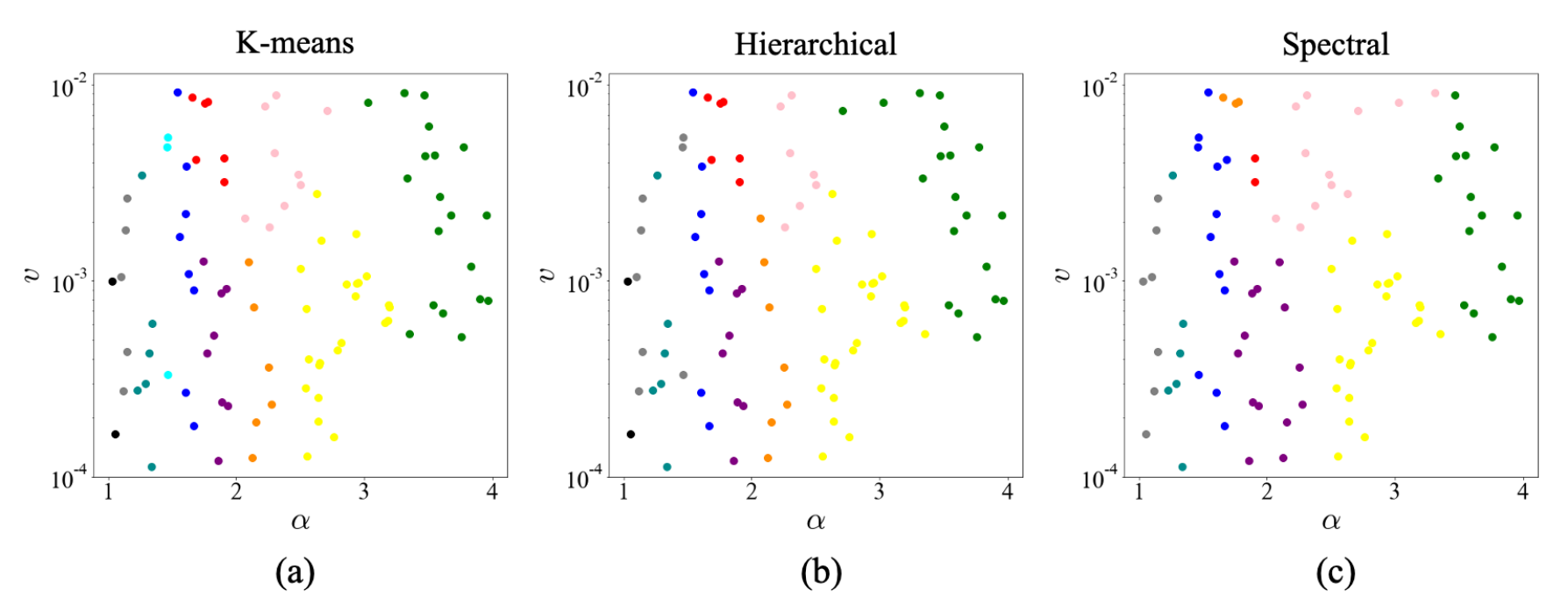}
  \caption{Clustering results from various clustering methods.
  (a) Clustering results using the K-means method.
  The number of clusters is determined using the elbow method~\cite{thorndike1953belongs} (see main text).
  (b) Clustering results using the hierarchical clustering method~\cite{maimon2005data}. 
  The hierarchy clustering method is adopted with the following hyperparameters: metric = euclidean, Linkage criteria = ward, and number of clusters K = 11. 
  (c) Clustering results using the spectral clustering method~\cite{von2007tutorial}. 
  The spectral clustering library is adopted with the following hyperparameters: affinity matrix = polynomial function with degree = 5, and kernel coefficient for polynomial function = 0.0001. 
  The number of clusters is optimized using the elbow method and determined as K = 9.}
  \label{fig_append5}
  \end{center}
\end{figure}

In the last analysis, K-means clustering results were compared with results determined by other clustering methods. 
As shown in Fig.~\ref{fig_append5}, the clustering results are similar to the K-means method results under appropriate parameter settings, 
which demonstrates the robustness of the clusters obtained by the K-means method.

\section{Procedure for quantification of intermediate states of PD}
\label{proc_inter}
\begin{figure}[htbt]
  \begin{center}
   \includegraphics[width=0.6\linewidth]{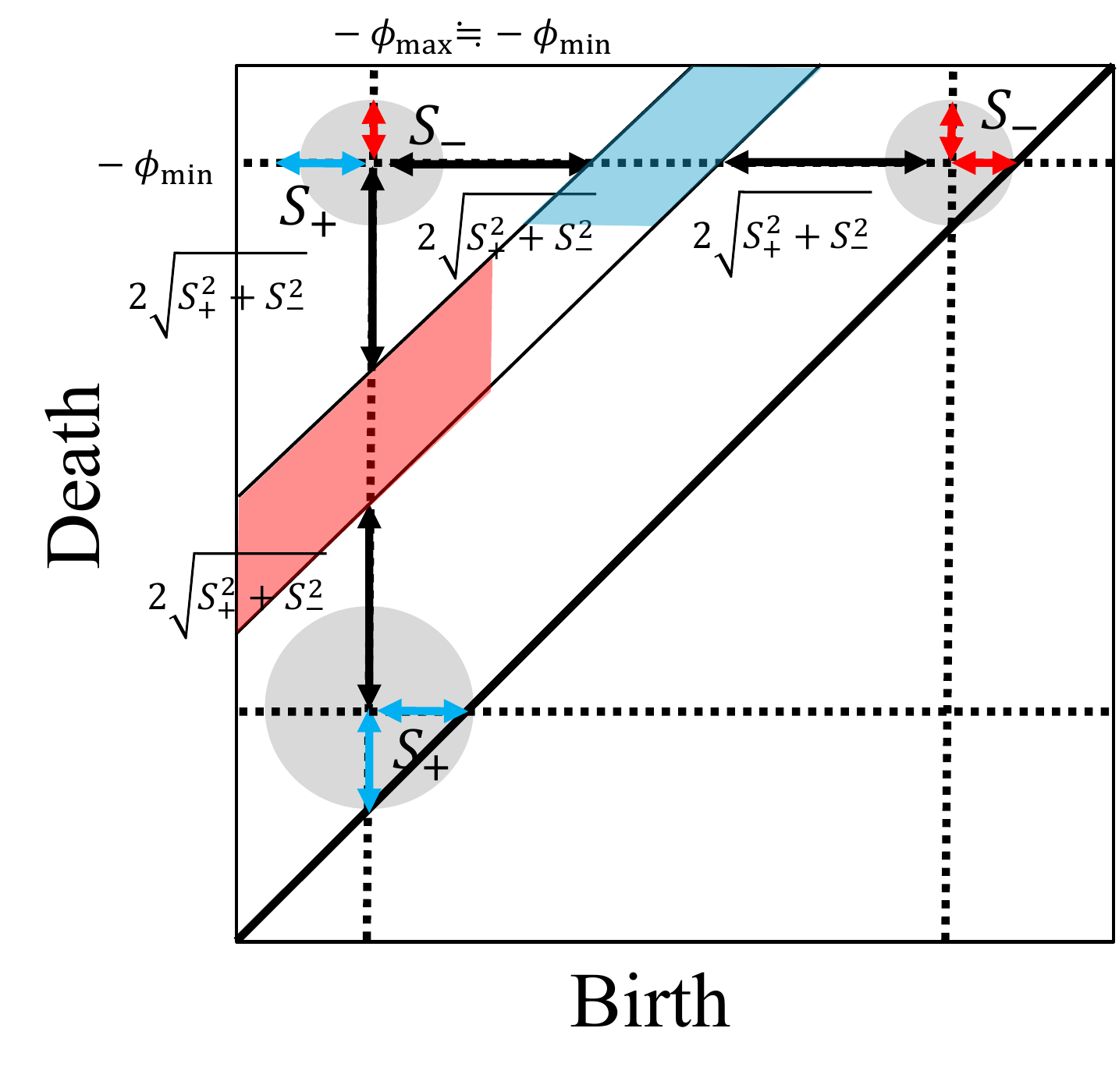}
  \caption{\textcolor{black}{%$N^{\rm gen}_{\rm mid}(t)$と$N^{\rm grow}_{\rm mid}(t)$を算出する際の、中間状態の領域の設定。
  Conceptual diagram of the intermediate state region setting when calculating $N^{\rm gen}_{\rm mid}(t)$ and $N^{\rm grow}_{\rm mid}(t)$.
  %赤い領域が$N^{\rm gen}_{\rm mid}(t)$に対応する中間領域で、青い領域が$N^{\rm grow}_{\rm mid}(t)$に対応する領域。
  The red region is the intermediate region corresponding to $N^{\rm gen}_{\rm mid}(t)$ and the blue region corresponds to $N^{\rm grow}_{\rm mid}(t)$. 
  %グレーの楕円と円がgeneratorのおおよその分布を表しており、この分布の広がりと、ヒストグラムの$S_{\rm +}$と$S_{\rm -}$が対応すると仮定して、領域を設定している。
  The gray ellipses and circles represent the approximate distribution of the generators, and the regions are set assuming that the width of this distribution corresponds to $S_{\rm +}$ and $S_{\rm -}$ in the histogram.}}
  \label{fig_area}
  \end{center}
\end{figure}

First, we generated a histogram from the magnetic domain pattern data $\phi\left(\bm{r}\right)_{t=2T_0}$ at $t=2T_0$, and the values of the positive peaks $\phi_{\rm max}$ and negative peak $\phi_{\rm min}$ were obtained. 
Similarly, we obtained thes standard deviations of the magnetic moment in the $\phi\left(\bm{r}\right)_{t=2T_0}\leq 0$ and $\phi\left(\bm{r}\right)_{t=2T_0}> 0$ regions called $\textcolor{black}{S}_{\rm +}$ and $\:\textcolor{black}{S}_{\rm -}$, respectively. 
We define the condition under which the lifetime $\Delta t'_{\rm mid\:life}$ of the hole in the intermediate state should be \textcolor{black}{the birth and death between $(-\phi_{\rm max} + 2S_{\rm +})$ and $(-\phi_{\rm min} - 2S_{\rm -})$.} 
\textcolor{black}{The generators with the following  lifetime $\Delta t'_{\rm mid\:life}$ are always satisfied under this condition (see Fig.~\ref{fig_area}).}
\textcolor{black}{
\begin{eqnarray}
\left(2\sqrt{S_{\rm +}^2 + \textcolor{black}{S}_{\rm -}^2}\right) < \Delta t'_{\rm mid\:life} < \left(\phi_{\rm max} - \phi_{\rm min} - 2\sqrt{S_{\rm +}^2 + S_{\rm -}^2}\right).
\end{eqnarray}
}
Note that the terms $t'$ and lifetime do not correspond to the time of the pattern formation process, but to the threshold for binarizing a grayscale image to calculate persistent homology. 
Moreover, the range of the birth time $t'_{\rm birth}$ values of the generators involved in domain generation is defined as
\begin{eqnarray}
t'_{\rm birth} \leq \textcolor{black}{-}\phi_{\rm \textcolor{black}{max}} + 2\textcolor{black}{S}_{\rm \textcolor{black}{+}},
\end{eqnarray}
and the range of the death time $t'_{\rm death}$ values of the generators involved in domain growth is defined as
\begin{eqnarray}
t'_{\rm death} \geq \textcolor{black}{-}\phi_{\rm \textcolor{black}{min}} - \textcolor{black}{S}_{\rm \textcolor{black}{-}}.
\end{eqnarray}
\textcolor{black}{These conditions were set because it was empirically confirmed that the generators of generation and growth can be discriminated by them.}

A hole that has a lifetime in the range of $\Delta t'_{\rm mid\:life}$ and its birth is earlier than $t'_{\rm birth}$ is defined as a generator in an intermediate state during domain generation. 
A hole that has a lifetime in the range of $\Delta t'_{\rm mid\:life}$ and its death is later than $t'_{\rm death}$ is defined as a generator in an intermediate state during domain growth. 
For example, from these definitions, the intermediate state of domain generation and domain growth corresponds to the red regions in Figs.~\ref{fig10}(a) and (c). 
Let $N^{\rm gen}_{\rm mid}$ and $N^{\rm grow}_{\rm mid}$ be the numbers of generators within these intermediate states of domain generation and growth, respectively, and let $N_{\rm all}$ be the number of generators in the PD. 
Then, the ratios of the number of generators in the intermediate states during domain generation and growth to the total number of generators are defined respectively as follows:
\begin{eqnarray}
R^{\rm gen}(t) &:=& \frac{N^{\rm gen}_{\rm mid}(t)}{N_{\rm all}(t)}\\
R^{\rm grow}(t) &:=& \frac{N^{\rm grow}_{\rm mid}(t)}{N_{\rm all}(t)}
\end{eqnarray}
As an indicator of whether or not intermediate states occur during the generation and growth, the maximum ratios of $R^{\rm gen}(t)$ and $R^{\rm grow}(t)$ are defined respectively as 
\begin{eqnarray}
R^{\rm gen}_{\rm max}&:=&R(t^{\rm gen}_{\rm max}),\:\:\:\:\:
t^{\rm gen}_{\rm max}:=\argmax_{t} R^{\rm gen}(t),\\
R^{\rm grow}_{\rm max}&:=&R(t^{\rm grow}_{\rm max}),\:\:\:\:\:
t^{\rm grow}_{\rm max}:=\argmax_{t} R^{\rm grow}(t).
\end{eqnarray}\par

\section{Analysis of the energy landscape in inverse magnetic domain generation}
\label{appendix2}

\begin{figure}[htbp]
  \begin{center}
   \includegraphics[width=\linewidth]{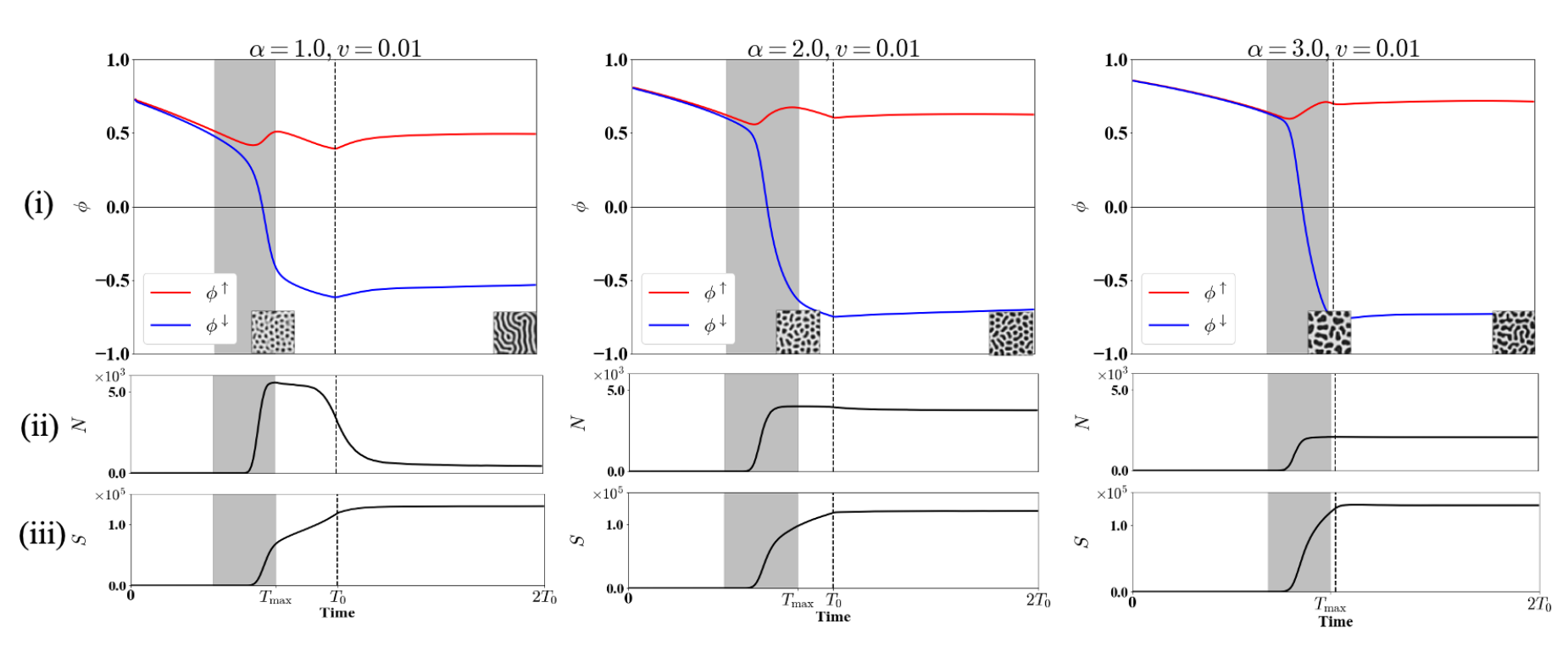}
  \caption{Time evolution of the average magnetic moment during domain generation at $v=0.01$ and $\alpha=1.0$, $2.0$, and $3.0$. 
  The time region of the domain generation is shown in gray. 
  The time at which the total number of domains $N$ is the largest is set to be the completion time of the domain generation $T_{\rm max}$, and the first time at which $N$ is greater than zero is set to be the start time of generation. 
  (i) Time evolution of the average magnetic moment $\phi^{\downarrow}$ in the region in which the magnetic moment becomes negative in column $T_{\rm max}$ and is maintained until $2T_0$ together with the time evolution of the average magnetic moment $\phi^{\uparrow}$ in the region in which it is positive. 
  (ii) Column time evolution of the total number of domains $N$. 
  (iii) Column time evolution of the area of the inverse magnetic domain region ($\phi < 0$).}
  \label{fig_append1}
  \end{center}
\end{figure}

\begin{figure}[htbp]
  \begin{center}
   \includegraphics[width=\linewidth]{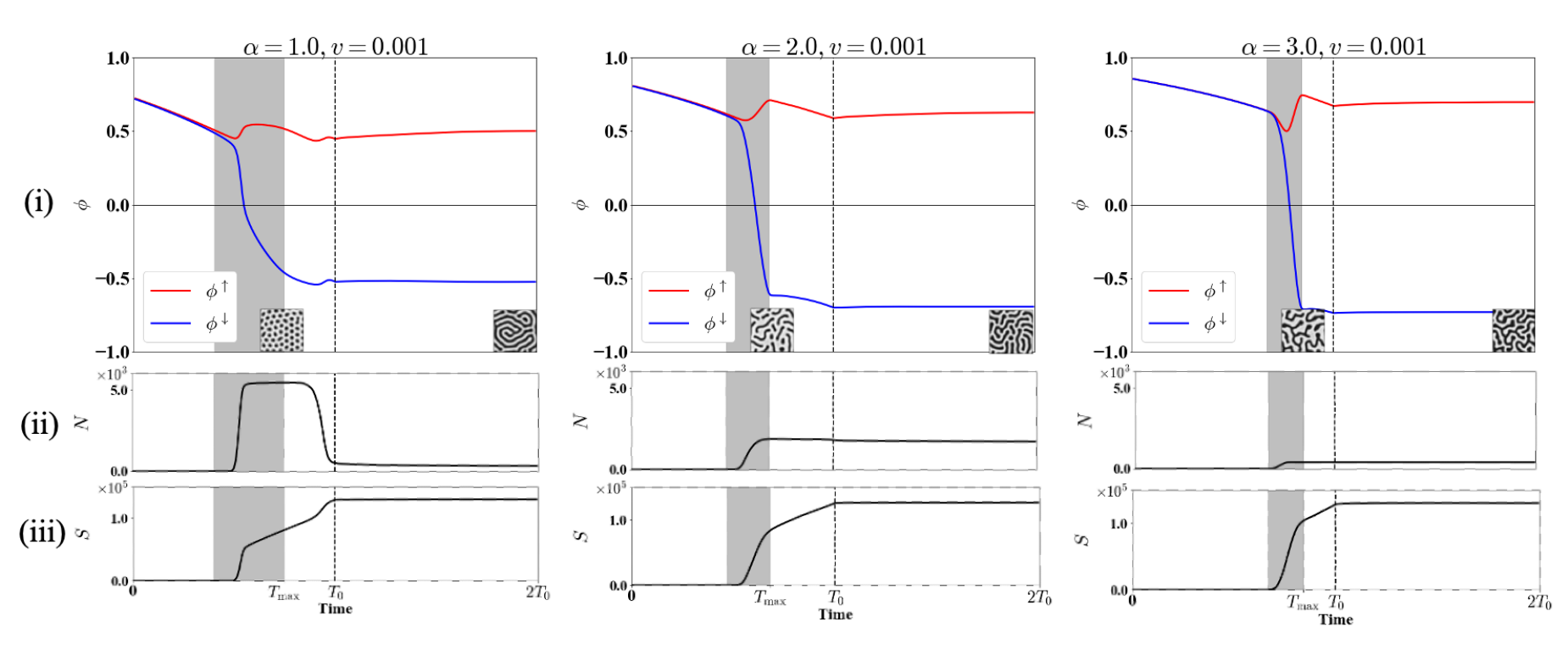}
  \caption{Time evolution of the average magnetic moment during domain generation at $v=0.001$ and $\alpha=1.0$, $2.0$, and $3.0$. 
  The time region of the domain generation is shown in gray. 
  The time at which the total number of domains $N$ is the largest is set to be the completion time of the domain generation $T_{\rm max}$, and the first time at which $N$ is greater than zero is set to be the start time of generation. 
  (i) Time evolution of the average magnetic moment $\phi^{\downarrow}$ in the region in which the magnetic moment becomes negative in column $T_{\rm max}$ and is maintained until $2T_0$ together with the time evolution of the average magnetic moment $\phi^{\uparrow}$ in the region in which it is positive. 
  (ii) Column time evolution of the total number of domains $N$. 
  (iii) Column time evolution of the area of the inverse magnetic domain region ($\phi < 0$).}
  \label{fig_append2}
  \end{center}
\end{figure}

In this appendix, by focusing on the structural change in the energy landscape, we analytically reveal the mechanism of the qualitative transition of the inverse magnetic domain generation process due to the increase in $\alpha$ as described in Sec.~\ref{sec_result2}. 
We define $T_{\rm max}$ as the time at which the number of isolated domains in the inverse magnetic domain reaches a maximum, and use it as the approximate end time of the inverse magnetic domain generation. 
The region that becomes an inverse magnetic domain at $T_{\rm max}$ and is maintained until $2T_0$ is denoted as $D_{T_{\rm max}}^{\downarrow}$, and the region that does not is denoted as $D_{T_{\rm max}}^{\uparrow}$. 
In the time domain in which the inverse magnetic domain occurs, the magnetic moment of $D_{T_{\rm max}}^{\downarrow}$ changes significantly, whereas that of $D_{T_{\rm max}}^{\uparrow}$ does not [Figs.~\ref{fig_append1}(i) and \ref{fig_append2}(i)]. 
Using this knowledge, the magnetic moment $\phi(\bm{r})$ at the time of domain generation can be simplified using representative values of a scalar variable $\phi$ as
\begin{eqnarray}
  \phi(\bm{r}) = 
  \begin{cases}
    \phi\:\:\: (\bm{r} \in D_{T_{\rm max}}^{\downarrow})& \\
    C(\alpha)\:\:\: (\bm{r} \in D_{T_{\rm max}}^{\uparrow})&
  \end{cases},
\end{eqnarray}
where $C={\rm const}$.\par
By substituting this simplified model into the energy function in Eq.~\eqref{energy}, we derive the coarse-grained energy function $H(\phi)$, expressed in terms of the representative values of the magnetic moments of the inverse magnetic domain, $\phi$. 
If the area of the inverse magnetic domain region at $T_{\rm max}$ is $S_{\rm max}$, the boundary length of the inverse magnetic domain region is $L_{\rm max}$, area of $D_{\downarrow}$ is $S$, and the boundary length is $L$. 
Thus, the energy function in Eq.~\eqref{energy} at the time of inverse magnetic domain generation is \textcolor{black}{approximated as}
\begin{eqnarray}
H(\phi) &=& \alpha S  \lambda \left( -\frac{\phi^2}{2} + \frac{\phi^4}{4} \right) \nonumber\\
&\:&+  \beta \frac{L(C(\alpha)-\phi)^2}{2}\nonumber\\
&\:&+\gamma \left[\int_{\bm{r} \in D_{T_{\rm max}}^{\downarrow}} d\bm{r} \int_{\bm{r}' \in D_{T_{\rm max}}^{\uparrow}} d\bm{r}' \frac{C(\alpha)\phi}{\left|\bm{r} - \bm{r}'\right|^3} + \int_{\bm{r} \in D_{T_{\rm max}}^{\downarrow}} d\bm{r} \int_{\bm{r}' \in D_{T_{\rm max}}^{\downarrow}} d\bm{r}' \frac{\phi^2}{\left|\bm{r} - \bm{r}'\right|^3}\right]\nonumber\\
&\:&-h(t)S\phi + {\rm const.},  \\
\label{landscape}
&=& w_0(\beta) + w_1(\beta,\gamma,h(t))\phi + w_2(\alpha,\beta,\gamma)\phi^2  + w_4(\alpha)\phi^4,
\end{eqnarray}
where $w_0$, $w_1$, $w_2$, and $w_4$ represent constants independent of $\phi$. 
If we consider only the case in which an isomorphic magnetic domain generates, we find that $w_1\propto -h(t) +C_1(\alpha)\propto t - C'_1(\alpha)$, $w_2\propto -\alpha + C_2$, $w_4\propto\alpha >0$, and $C_2 > 0$. 
Because only $w_1\phi$ is an asymmetric term, $w_1<0$ at $t=0$ changes to $w_1>0$ around the domain generation.\par
$\textcolor{black}{H}(\phi)$ has two minimal solutions when the discriminant formula of $\frac{d \textcolor{black}{H}(\phi)}{d \phi} = 0$,
\begin{eqnarray}
D=-\left(\frac{w_2(\alpha,\beta,\gamma)}{2w_4(\alpha)}\right)^3 - 27 \left(\frac{w_1(\alpha,\beta,\gamma)}{4w_4(\alpha)}\right)^2,
\end{eqnarray}
is positive, and one minimal solution when it is negative. 
Since $\frac{w_2}{w_4} \propto \frac{-\alpha + C_2}{\alpha}=-1+\frac{C_2}{\alpha}$ decreases as $\alpha$ increases, the discriminant expression, $D$, increases. 
In particular, if $\alpha$ is sufficiently large and the system exist around domain generation $w_1\sim 0$, then $w_2$ is a large negative value and $D>0$. 
This demonstrates that in a large region of $\alpha$, $\textcolor{black}{H}(\phi)$ has two minimal solutions around time of domain generation. 
Similarly, the fact that $w_1\propto t$ and $w_1>0$ after the domain generation suggests that the absolute value of $w_1$ increases with time, and the discriminant expression, $D$, decreases. 
This demonstrates that for an energy function with two minimal solutions around time of domain generation when the value of $\alpha$ is sufficiently large, one of the minimal states can be eliminated as time evolves. 
This change in the energy landscape explains the nucleation-like inversion of the magnetic domain at $\alpha>2.0$, because in the deterministic time-evolution model of the TDGL equation, the metastable state remains stable until it is resolved (Fig.~\ref{fig11}). 
Similarly, if the value of $\alpha$ is small and only one minimal solution exists even around the time of domain generation, the number of minimal solutions does not change with time. 
This explains the continuous domain generation at $\alpha<2.0$ (Fig.~\ref{fig11}). \par
From this reduced energy model, we can also discuss the mechanism of the sea-island state III and mixed states II and IV at around $\alpha = 2.0$. 
$\alpha \sim 2.0$ should be the boundary parameter region between the continuous and discontinuous magnetic domain generation ($w_2 \sim 0$). 
Therefore, owing to the spatial inhomogeneity $\lambda$ of anisotropy, some spatial regions are expected to be continuous and others discrete in magnetic domain generation. 
Compared with the case where magnetic domains are generated continuously, the discrete generation of magnetic domains is relatively delayed because the magnetic domains are not generated until the bimodal structure is resolved. 
A state in which there is already an inverse magnetic domain can be seen as an additional negative external magnetic field. 
This might prevent the resolution of the bimodal structure, thus creating a structure that partially lacks the labyrinthine structure. 
Although we only discussed the domain generation here, similar mechanisms are expected to work in domain growth. 
Thus, we can confirm that persistent homology is capable of extracting useful features that can reveal the mechanisms of the magnetic domain pattern formation process.\par

\section*{Acknowledgments}
This work was supported by JST, PRESTO Grant Number JPMJPR212A, JSPS KAKENHI Grant-in-Aid for Scientific Research on Innovative Areas ``Discrete Geometry for Exploring Next-Generation Materials'' 20H04648, 22K13979, 23H03460, 22H05106 and JST CREST JPMJCR15D3 and JPMJCR1861. We would like to thank Professor Satoshi Kuriki of the Institute of Statistical Mathematics for many helpful suggestions in the preparation of this paper. In addition, discussions with Senior Researcher Toshinobu Nakamura of the National Institute of Advanced Industrial Science and Technology provided deep insights into the advantages of persistent homology in physical data analysis.

%% If you have bibdatabase file and want bibtex to generate the
%% bibitems, please use
%%
 \bibliographystyle{elsarticle-num} 
 \bibliography{main}

%% else use the following coding to input the bibitems directly in the
%% TeX file.

% \begin{thebibliography}{00}

% %% \bibitem{label}
% %% Text of bibliographic item

% \bibitem{}

% \end{thebibliography}
\end{document}